\documentclass[aps,prd, nofootinbib,preprintnumbers,twocolumn]{revtex4}
\usepackage{amsmath,amsfonts,amssymb,graphicx,graphics,color,times,bbm,psfrag,dsfont,slashed}
\usepackage[latin9]{inputenc}
\usepackage{pifont}
\usepackage{natbib}
\usepackage{subfigure,epsfig,epstopdf}
\usepackage{bm}
\usepackage{enumerate}
\newcommand{\be}{\begin{equation}}
\newcommand{\ee}{\end{equation}}
\newcommand{\bea}{\begin{eqnarray}}
\newcommand{\eea}{\end{eqnarray}}

\begin{document}
 \def\be{\begin{equation}}
 \def\ee{\end{equation}}
 \def\l{\lambda}
 \def\a{\alpha}
 \def\b{\beta}
 \def\g{\gamma}
 \def\d{\delta}
 \def\e{\epsilon}
 \def\m{\mu}
 \def\n{\nu}
 \def\t{\tau}
 \def\p{\partial}
 \def\s{\sigma}
 \def\r{\rho}
 \def\sl{\ds}
 \def\ds#1{#1\kern-1ex\hbox{/}}
 \def\sla{\raise.15ex\hbox{$/$}\kern-.57em}
 \def\nn{\nonumber}
 \newcommand{\bth}{{\bf 3}}
 \newcommand{\btw}{{\bf 2}}
 \newcommand{\bon}{{\bf 1}}
 \def\Tr{\textnormal{Tr}}
 \def\th{\theta}
 \def\({\left(}
 \def\){\right)}
 \def\[{\left[}
 \def\]{\right]}

\preprint{}

\title{Intriguing aspects of meson condensation}
\author{Andrea Mammarella}
\email{andrea.mammarella@lngs.infn.it}
\author{Massimo Mannarelli}
\email{massimo@lngs.infn.it}
\affiliation{INFN, Laboratori Nazionali del Gran Sasso, Via G. Acitelli, 22, I-67100 Assergi (AQ), Italy}

\begin{abstract}
We analyze various aspects of pion and kaon condensation in the framework of chiral perturbation theory. Considering a system at vanishing temperature and varying  the isospin chemical potential and the strange quark chemical potential  we reproduce known results about the phase transition to the pion condensation phase and to the kaon condensation phase.   However,  we obtain  mesonic mixings and  masses  in the   condensed phases   that are in disagreement with the results reported in previous works. Our findings are obtained both  by a theory group analysis and by direct calculation by means of the same low-energy effective Lagrangian  used in previous works.
We also study the leptonic decay channels  in the normal phase and in the pion condensed phase, finding that some of these channels have a peculiar nonmonotonic behavior as a function of the isospin chemical potential. Regarding the  semileptonic decays, we find that that they are feeding processes for the  stable charged pion state.
\end{abstract}

\maketitle
\section{Introduction}
The properties of strongly interacting matter in an isospin and/or strangeness rich medium are relevant  in a wide range of phenomena including  the astrophysics of compact stars and heavy-ion collisions.  It is known that depending on the value of the isospin chemical potential, $\mu_I$, and on the value of the strangeness chemical potential,  $\mu_S$, three different phase can be realized: the normal phase, the pion condensed ($\pi c$) phase and the kaon condensed ($Kc$) phase~\cite{Migdal:1990vm,Son:2000xc,Kogut:2001id}. The realization of a mesonic condensate can drastically change the low energy properties of matter, including the mass spectrum and the lifetime of mesons.

Previous analyses of   the meson condensed phases by QCD-like theories were developed in \cite{Kogut:1999iv, Kogut:2000ek}. Pion condensation in two-flavor quark matter was studied in~\cite{Son:2000xc, Son:2000by} and in three-flavor quark matter in~\cite{Kogut:2001id}. In particular,  the phase diagram as a function of $\mu_I$ and $\mu_S$ was presented in~\cite{Kogut:2001id}. Finite temperature effects in  $SU(2)_L \times SU(2)_R$ chiral perturbation theory ($\chi$PT) have been studied in~\cite{Loewe:2002tw, Loewe:2004mu, He:2005nk, Xia:2014bla}. One remarkable property of quark matter with nonvanishing isospin chemical potential is that it is characterized by a real measure, thus the lattice realization can be performed with standard numerical algorithms~\cite{Alford:1998sd, Kogut:2002zg}. The $\pi c$ phase and the $K c$ phase have  been studied by NJL models in~\cite{Toublan:2003tt, Barducci:2004tt, Barducci:2004nc,Ebert:2005wr,Ebert:2005cs} and by random matrix models in \cite{Klein:2004hv}. All these models find results in qualitative and quantitative agreement, and in particular, the phase diagram of matter has been firmly established. However, regarding  the low energy mass spectrum in three-flavor quark matter, we found that it was only studied in~\cite{Kogut:2001id}. Our results are in disagreement with those of~\cite{Kogut:2001id}, the most relevant difference is in the  mixing between mesonic states.  Regarding the pion decay, previous works focused on density and temperature effects in standard decay channels~\cite{Barducci:1990sv, Dominguez:1993kr,Loewe:2011tm}, but not all the decay channels have been considered.

In the present paper we analyze the $\pi$c phase and the $K$c phase in a realization of  $\chi$PT~\cite{Gasser:1983yg, Leutwyler:1993iq, Ecker:1994gg,  Scherer:2002tk, Scherer:2005ri} that includes only the pseudoscalar mesons. Therefore,  the considered chiral Lagrangian approach is valid for  $|\mu_B|\lesssim940$  MeV, $|\mu_I|\lesssim 770$ MeV, and $|\mu_S|\lesssim550 $ MeV. These bounds come from the masses of the proton, the rho meson and the omega baryon, respectively. Moreover, $\chi$PT is valid in the energy range $E\lesssim 1$ GeV, corresponding to the breaking scale of the theory.
For definiteness we take the following values of the mesonic masses in  vacuum: $m_\pi = 140 $ MeV, $m_K= 495$ MeV and  $m_\eta= 547$ MeV. Unless explicitly stated, we will assume that in vacuum all the pion masses  and  all the kaon masses are equal. By this model  we discuss  the mixing and the masses of the pseudoscalar mesonic octet and the most relevant pion decay channels in the normal phase, in the $\pi$c phase and in the $K$c phase. Regarding the mesonic mixing, we discuss the disagreement with the results  of~\cite{Kogut:2001id} by theory group analysis and by explicit calculation using the $\chi$PT Lagrangian. Regarding the decay channels, since  the masses of the mesons strongly depend on $\mu_I$ and $\mu_S$, by changing these chemical potentials some decay channels can become kinematically forbidden  and/or other channels that are not allowed in vacuum can be opened.  

As we shall formally see, the presence of a baryonic chemical potential  is immaterial for the chiral Lagrangian, because mesons have no baryonic charge. However, it is clear that at large values of $\mu_B$ we expect a transition between hadronic matter and a different phase, presumably a color superconducting phase~\cite{Rajagopal:2000wf,Alford:2007xm,Anglani:2013gfu}. In principle we should limit ourselves to considering  $\mu_I < \mu_B$, however since the effective Lagrangian is blind to the baryonic chemical potential, we can assume that such inequality is always satisfied. Although we will consider the  range of values of $|\mu_I|\lesssim 770$ MeV, it is worth emphasizing that at asymptotic $\mu_I$ the system can be studied by perturbative QCD and the ground state is  a Fermi liquid with Cooper pairing of quarks~\cite{Son:2000xc, Son:2000by}.

One  interesting topic that to the best of our knowledge has not been previously  discussed in the pion and kaon condensed phases is the screening mass of the photon. By   the Nishijima-Nakano-Gell--Mann (NNG) formula
\be
Q= T_3 + \frac{Y}{2}\,,
\label{eq:NNG}
\ee
it is possible to relate the electric charge, $Q$, to  the third component of isospin, $T_3$, and  hypercharge, $Y$. In particular, if the vacuum carries isospin and/or strangeness charges, then it will be a superconductor because the $U(1)_Q$ gauge group will be broken. Thus, by the Higgs-Anderson mechanism the photon will acquire a Meissner mass. We evaluate the tree-level screening masses finding that in the two meson condesed phases  they have the same formal expression.   Moreover, the Debye and Meissner masses are equal.  In principle, any quark chemical potential breaks the Lorentz symmetry, therefore the Debye and Meissner masses of the photon can be different. However, we will show that  the tree-level  Lagrangian has to lead to equal Debye and Meissner masses.

The present paper is organized as follows. In Sec.~\ref{sec:general setting} we briefly review the  aspects of $\chi$PT  that are relevant for our work. In Sec.~\ref{sec:twoflavor} we consider  two-flavor quark matter. We discuss  pion condensation driven by an isospin chemical potential reviewing known results and generalizing the study of the low-energy Lagrangian.  In Sec.~\ref{sec:threeflavor} we consider three-flavor quark matter, determining the mixing angles and the masses of the pseudoscalar octet. In Sec.~\ref{sec:piondecays} we discuss the pion decay channels in the normal phase and in the $\pi c$ phase. In Sec.~\ref{sec:conclusion} we summarize our results. In the Appendix~\ref{appA} we discuss some details about the $\pi$-$W$  vertex factor relevant for pion decays.

\section{General setting}\label{sec:general setting}
In this section we briefly review the  aspects of chiral symmetry that are relevant for  meson condensation.   
The general ${\cal O}(p^2)$ Lorentz invariant Lagrangian density describing the pseudoscalar mesons can be written as 
\be\label{eq:Lagrangian_general}
{\cal L} = \frac{F_0^2}{4} \text{Tr} (D_\nu \Sigma D^\nu \Sigma^\dagger) + \frac{F_0^2}{4} \text{Tr} (X \Sigma^\dagger + \Sigma X^\dagger )\,,
\ee
where $\Sigma$ corresponds to the meson fields, $X=2 B_0( s + i p)$ describes scalar and pseudoscalar external fields and the covariant derivative is defined as 
\be
D_\mu \Sigma = \partial_\mu\Sigma - \frac{i}{2} [v_\mu,\Sigma] - \frac{i}{2} \{a_\mu,\Sigma \}\,,
\ee
with  $v_\mu$ and  $a_\mu$   the external vectorial and axial currents, respectively. The  Lagrangian has two free parameters $F_0$ and $B_0$, related to the pion decay and to the quark-antiquark condensate, respectively, see for example~\cite{Gasser:1983yg, Leutwyler:1993iq, Ecker:1994gg,  Scherer:2002tk, Scherer:2005ri}. 

The Lagrangian density is invariant under $SU(N_f)_L \times SU(N_f)_R $  provided the meson field transform as 
\be\label{eq:transSigma}
\Sigma \to R \Sigma L^\dagger\,,
\ee
and the chiral symmetry breaking corresponds to the spontaneous global symmetry breaking $SU(N_f)_L \times SU(N_f)_R \to SU(N_f)_{L+R}  $. The combination of the   $N_f^2-1$ Nambu-Goldstone bosons (NGBs), $\phi_a$ with $a=1, \dots, N_f^2-1$, corresponding to mass eigenstates can be   identified with the pseudoscalar  mesons fields. In standard $\chi$PT, the mass eigenstates are charge eigenstates as well. Thus  mesons are particles with a well defined mass  and charge. The presence of a medium can change this picture. In particular, if the vacuum carries  an electric charge, then the mass eigenstates will not typically be charge eigenstates. The presence of a medium can be taken into account by considering appropriate external currents in Eq.~\eqref{eq:Lagrangian_general}. 

At vanishing temperature the vacuum is determined by maximizing  the Lagrangian density with respect to the external currents. 
The pseudoscalar mesons are then described as  oscillations around the vacuum. We use the same nonlinear representation of~\cite{Kogut:2001id} corresponding to 
\be\label{eq:sigma}
\Sigma= u \bar \Sigma u \qquad \text{with} \qquad u=e^{i  T \cdot \phi/2} \,,
\ee
where $T_a$ are the $SU(N_f)$ generators and  $ \bar \Sigma $ is a generic $SU(N_f)$ matrix to be determined by maximizing the static Lagrangian. The reasoning behind the  above expression  is that under $SU(N_f)_L \times SU(N_f)_R $ mesons can be identified as   the fluctuations of the vacuum as in Eq.~\eqref{eq:transSigma} with $\theta_a^R = - \theta_a^L = \phi_a $.

In the following we will assume that $a_\mu=0$, $p=0$, $X=2 G M$, where $M$ is the $N_f \times N_f$ diagonal quark mass matrix and $G$ is a constant, that with these conventions is equal to $B_0$. 
Moreover, we will assume that $v^\nu = -2 e Q A^\nu - 2 \mu \delta^{\nu 0}$, meaning that the vectorial current consists of the electromagnetic field and a quark chemical potential, with $\mu$  a $SU(N_f) \times SU(N_f)$  matrix in flavor space. We first study the $N_f=2$  case and then the $N_f=3$ case. Since the two-flavor case is simpler to treat mathematically, it will allow us to establish a number of results that are useful for the description of the three-flavor case.

\section{Two-flavor case}\label{sec:twoflavor}
In two-flavor quark matter  the vacuum expectation value of the fields can be expressed as
\be\label{eq:barsigma}
\bar\Sigma = e^{i \bm\alpha\cdot\bm\sigma } = \cos \alpha + i \bm n\cdot\bm\sigma\sin \alpha\,,
\ee
where $ \bm\alpha=\bm n\alpha$ corresponds to the energetically favored direction in $SU(2)$ space. Assuming equal light quark masses, $m_u=m_d=m$, the  ${\cal O}(p^2)$ Lagrangian can be written as
\be\label{eq:Lagrangian}
{\cal L} = \frac{F_0^2}{4} \text{Tr} (D_\nu \Sigma D^\nu \Sigma^\dagger) + \frac{F_0^2 m_\pi^2}{2} \text{Tr} (\Sigma + \Sigma^\dagger )\,,
\ee
where $m_\pi^2 = 2 Gm/F_0^2$ is the pion mass for vanishing isospin chemical potential.  Expanding the covariant derivative we obtain 
\begin{align}\label{eq:Lagrangian2}
{\cal L} =& \frac{F_0^2}{4} \text{Tr} (\partial_\nu \Sigma \partial^\nu \Sigma^\dagger) + \frac{F_0^2 m_\pi^2}{2} \text{Tr} (\Sigma + \Sigma^\dagger ) \nonumber\\&-\frac{F_0^2}{16} \text{Tr} [v^\mu,\Sigma][v_\mu,\Sigma^\dagger] - \frac{i F_0^2}{4}  \text{Tr } \partial^\mu \Sigma  [v_\mu, \Sigma] \,,
\end{align}
and considering   the quark chemical potential
\be \mu={\rm diag}(\mu_u,\mu_d) = \frac{\mu_B}{3}+ \frac{\mu_I \sigma_3}{2}\,,\ee
we can write
\be
v^\nu = -2 e Q A^\nu - 2 \mu \delta^{\nu 0} = - \tilde A^\nu_I I - \tilde A^\nu_3 \sigma_3\,,
\ee
with
\begin{align}
\tilde A^\nu_I = \frac{1}{3} (eA_0 +\mu_B,e\bm A)\,, \\ 
\tilde A^\nu_3 =  (eA_0 +\mu_I,e\bm A)\label{eq:Atilde}\,. 
\end{align}
Given that in Eq.~\eqref{eq:Lagrangian2} the interaction terms between $v_\mu$ and $\Sigma$ are proportional to commutators of these two fields,  the only relevant term in $v_\mu$ is the one proportional to  $\tilde A^\mu_3$, and this is consistent with the fact that mesons have no baryonic charge.
Note that both $\mu_I$ and $M$ explicitly break $SU(2)_L \times SU(2)_R$  chiral symmetry giving mass to the (pseudo) NGBs. Equal light quark masses leave $SU(2)_{I}  $ invariant  ensuring that  pions  have equal masses. The isospin chemical potential induces a further symmetry breaking, such that $SU(2)_I  \to U(1)_{L+R}$ with the effect of removing the pion mass degeneracy with a  contribution  proportional to the isospin charge. Since pions are an isotriplet, it follows that the contribution of the isospin chemical potential to  the $\pi_0$ mass vanishes and the contributions to the $\pi_\pm$ is a Zeeman-like splitting, thus 
\begin{align}
    m_{\pi^0} &= m_\pi\,, \\
      m_{\pi^\pm}&= m_\pi \mp \mu_I\,;
\end{align}
clearly the   condensation of charged pions happens at $|\mu_I| = m_\pi$. The  only symmetry of the Lagrangian in Eq.~\eqref{eq:Lagrangian} is $U(1)_{L+R} $; when it is spontaneously broken it leads to a massless NGB, corresponding to one of the two charged pions depending on the sign of the isospin chemical potential. 

At the microscopic level, the breaking pattern induced by the isospin chemical potential and the light quark masses is
\be
\underbrace{SU(2)_L \times SU(2)_R\times U(1)_B }_{\displaystyle\supset[U(1)_Q]}
\to \underbrace{U(1)_{L+R}\times  U(1)_{B}}_{\displaystyle\supset [U(1)_Q]}\,, 
\ee 
where $U(1)_Q$ corresponds to the electromagnetic gauge symmetry.  In the broken phase one of the two charged pions condense, spontaneously breaking the $U(1)_Q$ symmetry, meaning that   the system  becomes an electromagnetic superconductor. Formally,  $Q$ can be expressed as a combination of the generator of $U(1)_B$ and of $U(1)_{L+R}$, thus the breaking of  $U(1)_{L+R}$ leads to a screening mass for the photon by the Higgs-Anderson mechanism.

Regarding the Lorentz symmetry, the isospin chemical potential explicitly breaks boost symmetry, however by expressing the isospin chemical potential as the expectation value of the $\tilde A^\mu_3 $ field we can formally consider a Lorentz invariant Lagrangian. To formally preserve Lorentz symmetry we will as well employ the Lorenz gauge $\partial_\mu \tilde A^\mu_3 =0$.

\subsection{Ground state}
For vanishing mesonic fluctuations the Lagrangian is a functional of $\bar \Sigma$ and $\tilde A^\mu_3$;  upon substituting Eq.~\eqref{eq:barsigma} in Eq.~\eqref{eq:Lagrangian} we obtain
\be\label{eq:L0_tilde}
{\cal L}_0(\alpha,n_3,\tilde A^\mu) = F_0^2 m_\pi^2 \cos\alpha + \frac{F_0^2}2 \sin^2\alpha\tilde A^\mu_{3}\tilde A_{3 \mu}(1-n_3^2)\,,\ee
which is a function of the parameters $\alpha$ and $n_3$ and a functional of $\tilde A^\mu_3$. For vanishing external electromagnetic field  and for $\mu_I < m_\pi$, the global maximum is at $\cos\alpha=1$ and ${\cal L}_0$ is independent of $\bm n$, meaning that the ground state has an $SU(2)$ global symmetry. In this case the custodial $SU(2)$ is still present and only the curvature of the potential (the pion masses) are affected by the isospin chemical potential. In other words, the isospin chemical potential is not sufficient to tilt the vacuum in one direction, thus  the  vacuum is the same obtained with $\mu_I=0$. 

The stationary point of ${\cal L}_0(\alpha,n_3,\delta_{\mu 0} \mu_I) $ corresponds to   $n_3=0$ and $\cos\alpha_\pi=m_\pi^2/\mu_I^2$, which  is a global maximum for $\mu_I > m_\pi$. In this case the vacuum is tilted by an angle $\alpha_\pi = \arccos (m_\pi^2/\mu_I^2) $ and the ground state has only a residual $O(2)$ symmetry (isomorphic to $U(1)$) for rotations $n_1=\cos\theta$ and $n_2=\sin\theta$;  the angle $\theta$ cannot  be determined maximizing the ground state Lagrangian and is signaling the existence of a massless NGB. 

The ground state Lagrangian can be easily determined and is given by
\be\label{eq:barL0}
{\cal \bar L}_0 = \left\{ \begin{array}{lr} F_0^2 m_\pi^2 & \text{for  } \mu_I < m_\pi \\ \frac{1}{2}F_0^2\mu_I^2 \left(1+\frac{m_\pi^4}{\mu_I^4}\right) & \text{for  } \mu_I > m_\pi \end{array}\right.\,.
\ee

Regarding the screening masses of the electromagnetic field, they can be inferred from Eq.~\eqref{eq:L0_tilde}.  The electromagnetic field has both a Debye mass and a Meissner mass, which are equal and given by 
\be
M_D^2 = M_M^2 = F_0^2 e^2 (\sin\alpha)^2 \,.
\ee
The screening masses vanish in the unbroken phase and are equal to  $F_0^2 e^2 (1-m_\pi^2/\mu_I^2)^2$ in the broken phase, signaling the breaking of $U(1)_Q$.  In principle,  the Debye and Meissner masses could be different, because the Lorentz symmetries is explicitly broken  by $\mu_I$. However, from the fact that the isospin chemical potential can be introduced as in Eq.~\eqref{eq:Atilde} it is clear that both tree-level screening masses must be equal.

\subsubsection{Generic chemical potential}
To properly understand the previous results regarding the ground state configuration we  consider  a more  general setting with
\be
\mu = \frac{1}{2} \bm \mu \cdot \bm \sigma\,,
\ee
corresponding to a quark chemical potential pointing to an arbitrary direction in isospin space.
The ground state Lagrangian is obtained maximizing
\be
{\cal L}_0 = \frac{F_0^2}{2} (\sin\alpha)^2  (|\bm\mu|^2-|\bm\mu\cdot \bm n|^2) + F_0^2m_\pi^2 \cos\alpha\,, 
\ee
as a function of $\alpha$ and $\bm n$. It is clear that $\bm\mu \perp \bm n$, thus  $\bm n$ is in the plane perpendicular to $\bm\mu$. This leads to the  residual $O(2)$ symmetry for rotations around $\bm\mu$. For  $ |\bm\mu| > m_\pi$ the ground state is tilted by an  angle  $\alpha_\pi=\arccos(m_\pi^2/|\bm\mu|^2)$.  The ground state Lagrangian  is the same reported in Eq.~\eqref{eq:barL0}, but with $\mu_I \to |\bm\mu|$.

\subsection{Quadratic Lagrangian \label{quadL}}
The leading order Lagrangian describing the in medium pions can be obtained expanding Eq.~\eqref{eq:Lagrangian2} at the second order in the fields. For definiteness we consider  $\bm \mu = (0,0,\mu_I)$ and $\bm n = (n_1, n_2, 0) = (\cos\theta, \sin\theta,0)$ in the vev in Eq.~\eqref{eq:barsigma}.
We decompose the ${\cal O}(p^2)$ Lagrangian at the second order in the fields as follows
\be
{\cal L}_\text{eff} = {\cal L}_K + {\cal L}_M + {\cal L}_L\,,
\ee
where ${\cal L}_K$ is the kinetic term, ${\cal L}_M $ is the mass term and ${\cal L}_L$ is the term linear in the derivatives. 

The kinetic part of the Lagrangian can be written as
\begin{widetext}
\be\label{eq:kinsu2}
 {\cal L}_K  = \frac{F_0^2}{2} \left( \delta_{ab}(\cos\alpha)^2 +   n_an_b (\sin\alpha)^2\right)
\partial_\nu \phi_a \partial^\nu \phi_b = \frac{F_0^2}{2} \partial_\nu \phi_a K_{ab} \partial^\nu  \phi_b\,,
\ee
\end{widetext}
that manifestly shows meson mixing. Since  $K$ is a symmetric matrix, it can be diagonalized.  By the transformation  
\begin{align}
\phi_1 &= \frac{1}{F_0}\left(n_1 \tilde \phi_1  - \frac{n_2 \tilde\phi_2}{\cos\alpha}\right)\,, \label{eq:phi1}\\
\phi_2 &= \frac{1}{F_0}\left(\frac{n_1 \tilde \phi_2}{\cos\alpha}  + n_2 \tilde\phi_1\right)\,,\label{eq:phi2} \\
\phi_3 &= \frac{\tilde\phi_3}{F_0\cos\alpha} \,,
\end{align}
we obtain  the canonical kinetic term
\be\label{eq:kinsu2f}
 {\cal L}_K  = \frac{1}{2}\partial_\nu \tilde\phi_a \partial_\nu \tilde\phi_a \,.
\ee
One of the peculiar aspects of the field redefinition above is that in the $\pi c$ phase for $m_\pi/\mu_I \to 0$ the terms proportional to $(\cos\alpha_\pi)^{-1}$ diverge. In other words, for vanishing light quark masses the  above field renormalization does not seem to work. The correct prescription for handling this issue  seems to be  to consider the $m_\pi/\mu_I\to 0$ limit only in the physical results. 

Regarding the electric charge eigenstates, we find that
\be
\pi^{\mp} = \frac{e^{\pm i \theta}}{F_0 \sqrt{2}} \left(\tilde\phi_1 \pm i \frac{\tilde\phi_2}{\cos\alpha}\right)\,,
\ee
where $e^{\pm i \theta} = n_1 \pm i n_2$. Note that the standard definition of the charge eigenstates is obtained for $\cos\alpha=1$, as expected.

For the mass term we find 
\begin{align}
{\cal L}_M =&  -\frac{m_\pi^2}{2} F_0^2  \cos\alpha\phi_a \phi_a \nonumber \\ &+ \frac{F_0^2}{2} \tilde A_e^\mu\tilde A_{3\mu}[\cos^2\alpha (\phi_1^2+\phi_2^2) - \sin^2\alpha(\bm n \cdot \bm \phi)^2]\,,
\end{align}
that in the rotated basis turns out to be
\begin{align}
{\cal L}_M =&  -\frac{m_\pi^2}{2} \cos\alpha \left(\tilde\phi_1^2 +\frac{\tilde\phi_2^2+\tilde\phi_3^2}{\cos^2\alpha}\right)  \\ &+ \frac{\tilde A_3^\mu\tilde A_{3\mu}}{2} \left[\tilde\phi_1^2 (\cos^2\alpha-\sin^2\alpha) +  \tilde\phi_2^2\right]\,. \nonumber
\end{align}

The term with a linear dependence on the  derivative is given by 
\be\label{eq:linear}
{\cal L}_L =  \frac{i F_0^2}{2}  \text{Tr}  \tilde A^\mu_3 [\Sigma^\dagger, \partial_\mu\Sigma] = -2 F_0^2  \tilde A^\mu_3\phi_1\partial_\mu\phi_2 \cos^2\alpha\,,
\ee
that by the rotated basis redefinition turns in
\be\label{eq:linear_tilde}
{\cal L}_\text{L} = -2  \tilde A^\mu_3\tilde\phi_1\partial_\mu{\tilde{\phi}}_2 \cos\alpha\,.
\ee
An interesting aspect is that in the $\pi c$ phase this is the only mixing term between the $\tilde\phi$ fields. Since it scales as $\cos\alpha_\pi$, it vanishes for $m_\pi/\mu_I \to 0$. 

Note that no term  of the quadratic Lagrangian depends on $\bm n = (n_1, n_2, 0)$, thus the ${\cal O}(2)$ symmetry has been absorbed in the redefinition of the fields. Assuming that no electromagnetic field is present, we can replace in all the Lagrangian terms $ \tilde A^\mu_3 \to \delta^{\mu 0} \mu_I$, obtaining in momentum space
\begin{widetext}
\begin{align}
{\cal L}_\text{eff} &= (\tilde\phi_1\,  \tilde\phi_2\,  \tilde\phi_3)\left(\begin{array}{ccc} k^2-m_\pi^2 \cos\alpha+\mu_I^2\cos(2\alpha) &- 2 i k_0 \mu_I \cos\alpha & 0\\ 2 i  k_0 \mu_I \cos\alpha &   k^2-m_\pi^2/\cos\alpha+\mu_I^2& 0\\ 0 & 0&  k^2-m_\pi^2/\cos\alpha\\ \end{array}\right)\left(\begin{array}{c} \tilde\phi_1 \\ \tilde\phi_2 \\ \tilde\phi_3\end{array}\right)\nonumber \\ &= \tilde \Phi S^{-1}  \tilde \Phi^t\,,
\end{align}
\end{widetext}
where $k^2=k_0^2-\bm k^2$ and $S^{-1}$ is the inverse propagator. 
The energy spectrum is obtained from the poles of the propagator and in the $\pi c$ phase we find   
\begin{widetext}
\begin{align}
E_{\pi_0} &= \sqrt{p^2+\mu_I^2}\,, \label{eq:pi0}\\
E_{\tilde\pi_+} &= \frac{1}{\sqrt{2} \mu_I}\sqrt{3 m_\pi^4 + \mu_I^4 + 2 p^2 \mu_I^2 -\sqrt{(3 m_\pi^4 + \mu_I^4)^2+16 m_\pi^4\mu_I^2 p^2}}= p \sqrt{\frac{\mu_I^4-m_\pi^4}{3 m_\pi^4 + \mu_I^4}} + {\cal O}(p^2)\,, \\
E_{\tilde\pi_-} &=\frac{1}{\sqrt{2} \mu_I}\sqrt{3 m_\pi^4 + \mu_I^4 + 2 p^2 \mu_I^2 +\sqrt{(3 m_\pi^4 + \mu_I^4)^2+16 m_\pi^4\mu_I^2 p^2}}= \frac{\sqrt{3 m_\pi^4 +\mu_I^4}}{\mu_I} + \frac{\mu_I p^2}{2} \sqrt{\frac{7 m_\pi^4 + \mu_I^4}{(3 m_\pi^4 + \mu_I^4)^{3/2}}} + {\cal O}(p^4) \label{eq:Etildepimeno} \,,
\end{align}
\end{widetext}
where $\tilde\pi_\pm$ are the two mass eigenstates (note that in this case the subscript does not indicate the electric charge). The dispersion law of the massless mode $\tilde\pi_+$ is linear in momentum with a  velocity that tends to the speed of light for $m_\pi/\mu_I \to 0$ and that vanishes for $m_\pi/\mu_I \to 1$. The $\pi_\pm$  charge eigenstates can be expressed as   a linear combination of the $\tilde\pi_\pm$ fields  as follows: 
\begin{widetext}
\be\label{eq:pimixing}
\pi_{\mp} = \frac{ie^{\pm i \theta}}{ (a_+-a_-)\sqrt{2}} \left[\sqrt{1+a_-^2} \left(1\pm \frac{a_+}{\cos\alpha}\right)\tilde\pi_- -\sqrt{1+a_+^2} \left(1\pm \frac{a_-}{\cos\alpha}\right) \tilde\pi_+  \right]\,,
\ee
\end{widetext}
where 
\be a_\pm= \frac{\mu_I^4-m_\pi^4\pm \sqrt{(\mu_I^4-m_\pi^4)^2+16  k_0^2 m_\pi^4 \mu_I^2 } }{4  k_0 m_\pi^2 \mu_I}\,.
\ee
Given the particular expression of the $a_\pm$ coefficients, the propagating particles oscillate between the two electric charge  eigenstates  with a  mixing angle depending on the energy.  This is a rather peculiar behavior because in the Standard Model one typically has  mixing angles that are not energy/momentum dependent. Note that this oscillation also means that the propagating particles oscillate between isospin eigenstates. 
This is possible because in the condensed phase the vacuum carries isospin charge which is related to the electric charge by the NNG formula \eqref{eq:NNG}, thus nor the  electric charge nor the isospin charge are  conserved. 

Eq.~\eqref{eq:pimixing}  can be inverted to obtain
\be
\left( \begin{array}{c}  \tilde\pi_+ \\ \tilde\pi_- \end{array} \right) = \left(\begin{array}{cc} U_{11} & U_{12} \\ U_{21} & U_{22}\end{array}  \right ) \left( \begin{array}{c}  \pi_+ \\ \pi_- \end{array} \right)\,, 
\label{eq:inverted}
\ee
and defining 
\be
s_{12}=\frac{2 m_\pi^2}{\mu_I}  \qquad M_2^2=-\frac{m_\pi^4-\mu_I^4}{\mu_I^2}\,, 
\ee
we find that 
\be
U=\left( \begin{array}{cc}
\frac{1}{n_-} & \frac{1}{n_-} \frac{M_2^2-\sqrt{M_2^4+4 k_0^2 s_{12}}}{2 i k_0 s_{12}} \\
\frac{1}{n_+} & \frac{1}{n_+} \frac{M_2^2+\sqrt{M_2^4+4 k_0^2 s_{12}}}{2 i k_0 s_{12}}
\end{array} \right)\,,
\label{U}
\ee
with 
\be
n_{\pm}=\frac{8 k_0^2 s_{12}^2+2 M_2^4\mp 2 M_2^2\sqrt{M_2^4+4 k_0^2 s_{12}^2}}{4 k_0^2 s_{12}^2}\,.
\ee

This result is important for the determination of the width of the pion decays discussed in Sec.~\ref{sec:piondecays}.

The mixing between the charged pion states can be simply understood in two-flavor quark matter. The $\pi_\pm$ states are the only states having a non vanishing value of the third component of isospin, and since the vacuum has a nonvanishing $\mu_I$, these states can mix. In the three-flavor case things become a little more involved. 

\section{Three-flavor case}\label{sec:threeflavor}
In three-flavor quark matter besides the isospin chemical potential one has to consider the strange quark chemical potential. Microscopically, strange quark states can be occupied by electroweak processes if the light quark chemical potential exceeds the strange quark mass. The formal expression of the in medium effective chiral Lagrangian is given by  Eq.~\eqref{eq:Lagrangian} in which  the mesonic octet is introduced by replacing
\be
u=e^{i  \phi_a\lambda_a/2 } \label{eq:u_three}\,,
\ee  
in Eq.~\eqref{eq:sigma}, where $ \lambda_a$ are the Gell-Mann matrices.

The isospin and strange quark chemical potential can be introduced by considering
\begin{widetext}
\be
\mu=\text{diag}\left(\mu_u,\mu_d,\mu_s\right)= \text{diag}\left(\frac13 \mu_B+\frac12 \mu_I,\frac13 \mu_B-\frac12 \mu_I, \frac13
\mu_B-\mu_S\right)=\frac{\mu_B-\mu_S}3 I + \frac{\mu_I}2 \lambda_3 + \frac{\mu_S}{\sqrt{3}} \lambda_8\,,
\ee
\end{widetext}

where $\mu_S$ is the so-called strange quark chemical potential. Note that the actual strange quark chemical potential is 
$\mu_s=\frac13 \mu_B-\mu_S$, however the diagonal contribution of the baryonic chemical potential  is immaterial for mesons. 

For three-flavor quark matter the spontaneous symmetry breaking pattern is the following
\be
\underbrace{SU(3)_L \times SU(3)_R }_{\displaystyle\supset[U(1)_Q]}
\to \underbrace{SU(3)_{V}}_{\displaystyle\supset [U(1)_Q]}\,, 
\ee 
and the corresponding $8$ NGBs are identified with the mesonic pseudoscalar octet. The quark masses explicitly break the chiral symmetry, giving mass to the pseudo NGBs. A similar effect is produced by the isospin chemical potential and the strange quark chemical potential, with the additional fact of breaking Lorentz symmetry. The symmetry of  the Lagrangian is thus reduced to 
\be
\underbrace{U(1)_{L+R} \times U(1)_{L+R}}_{\displaystyle\supset [U(1)_Q]} \,.
\ee
The  breaking of this symmetry leads, at most, to the appearance of one NGB and of the screening masses for  the electromagnetic field.

For the external vector current we can write
\be
v^\nu = -2 e Q A^\nu - 2 \mu \delta^{\nu 0} = - \frac{2}3 (\mu_B-\mu_S) I \delta^{\nu 0}- \tilde A^\nu_3 \lambda_3 - \tilde A^\nu_8 \lambda_8\,,
\ee
where
\begin{align}
\tilde A^\mu_3 &= (eA_0 +\mu_I,e\bm A)\label{eq:Atilde3_3flavor}\,, \\ 
\tilde A^\mu_8 &=  (eA_0 +2 \mu_S,e\bm A)\label{eq:Atilde8_3flavor}\,, 
\end{align}
are the relevant components of the electromagnetic  field. 

\subsection{Ground state}
In the three-flavor case the most general vev,  $\bar{\Sigma}$, depends on $8$ parameters, corresponding to the possible orientations in $SU(3)$ space. However,  in the two-flavor case we have found that  rotations around the direction of the chemical potential leave the vacuum invariant. We assume that the   same is true  in the three-flavor case and therefore the vacuum Lagrangian only depends on two angles, $\alpha$ and $\beta$, corresponding to the angles between the vacuum and third component of isospin, and  between the isospin and the hypercharge, respectively. This is exactly the same assumption used in~\cite{Kogut:2001id}, in which it was found that there are three different vacua:
\begin{itemize}
\item Normal phase: \begin{align} \mu_I&<m_\pi\,,\\ \mu_S&<m_K-\frac12 \mu_I\,,\end{align}
 characterized by
\be
     \alpha_N=0,  \quad\beta_N \in (0,\pi), \quad \bar{\Sigma}_N=\text{diag}(1,1,1)\,.
\ee
\item Pion condensation phase: \begin{align} \mu_I&>m_\pi\,, \\ \mu_S&<
\frac{-m_\pi^2+\sqrt{(m_\pi^2-\mu_I^2)^2+4 m_K^2 \mu_I^2} }{2 \mu_I}\,,
\end{align}
 characterized by
\be
    \cos \alpha_{\pi}=\left(\frac{m_\pi}{\mu_I}\right)^2,  \quad \beta_{\pi} =0\,,\ee  \begin{align} \bar{\Sigma}_{\pi}&=\left( \begin{array}{ccc}
                            \cos \alpha_\pi & \sin \alpha_\pi & 0 \\
                            -\sin \alpha_\pi & \cos \alpha_\pi & 0\\
                            0 & 0 & 1
                            \end{array} \right) \\ &= \frac{1+2 \cos\alpha_\pi}{3} I + i \lambda_2 \sin\alpha_\pi+ \frac{\cos\alpha_\pi-1}{\sqrt{3}}\lambda_8\,. \nonumber
                            \label{pion}
\end{align}
\item Kaon condensation phase: 
\begin{align}
\mu_S&>m_K-\frac12 \mu_I\,, \\ 
\mu_S&>
\frac{-m_\pi^2+\sqrt{(m_\pi^2-\mu_I^2)^2+4 m_K^2 \mu_I^2 }}{2 \mu_I}\,,
\end{align}
characterized by
\be
\cos \alpha_K=\left( \frac{m_K}{\frac12 \mu_I+\mu_S}\right)^2\,,  \beta_K=\pi/2\,,\ee 
\begin{align} \label{kaon} \bar{\Sigma}_K
=&\left( \begin{array}{ccc}
                            \cos \alpha & 0 & \sin \alpha  \\
                            0 & 1 & 0\\
                            -\sin \alpha & 0 & \cos \alpha
                            \end{array} \right)=\\
&\frac{1+2 \cos\alpha_K}{3} I+ \frac{\cos\alpha_K-1}{2\sqrt{3}}\left(\sqrt{3}\lambda_3-\lambda_8\right) + i \lambda_5 \sin\alpha_K\,. \nonumber 
\end{align}
Note that the kaon condensation  can only happen for
\be \mu_S > \bar\mu_S=m_K - \frac{m_\pi}2\,,\ee 
and  $ \bar\mu_S= 425$ MeV for our parameter choice.
\end{itemize}

\subsubsection{Screening masses}
As in the two-flavor case we can determine the screening masses of the electromagnetic field. Remarkably, we find that the screening masses are independent of the $\beta$ angle and have the  same  expression obtained in the two-flavor case
\be\label{eq:screening}
M_D^2 = M_M^2 = F_0^2 e^2 (\sin\alpha)^2 \,.
\ee
The nonvanishing value of the Meissner mass implies that  in both mesonic condensed phase the system is an electromagnetic superconductor. Note that across the first order phase transition between the two condensed phases the screening masses are discontinuous, because  $\alpha$ is discontinuous.

\subsection{Mixing}\label{sec:mixing}
\begin{figure}[th!]
\includegraphics[width=8.cm]{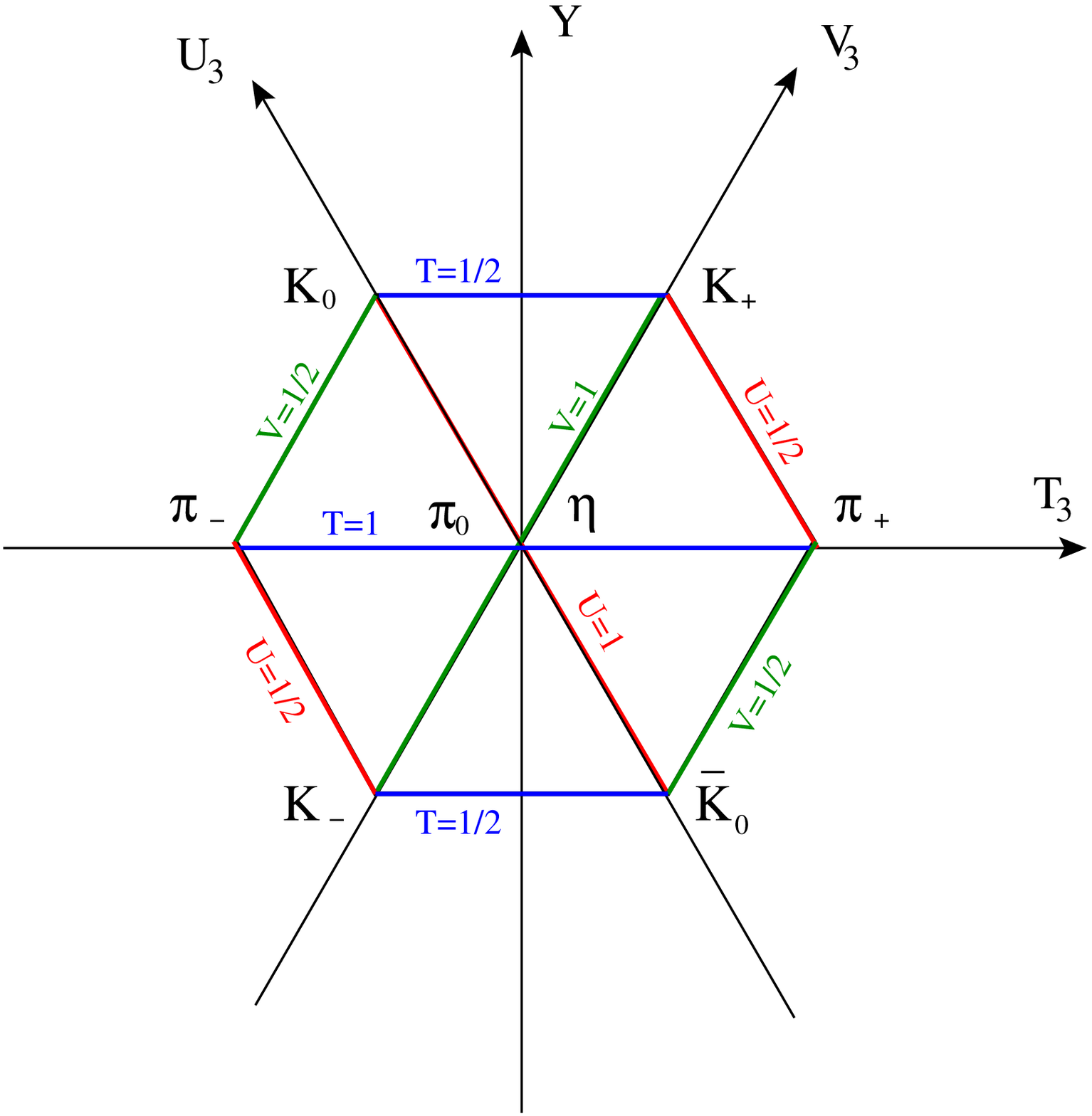}
\includegraphics[width=8.cm]{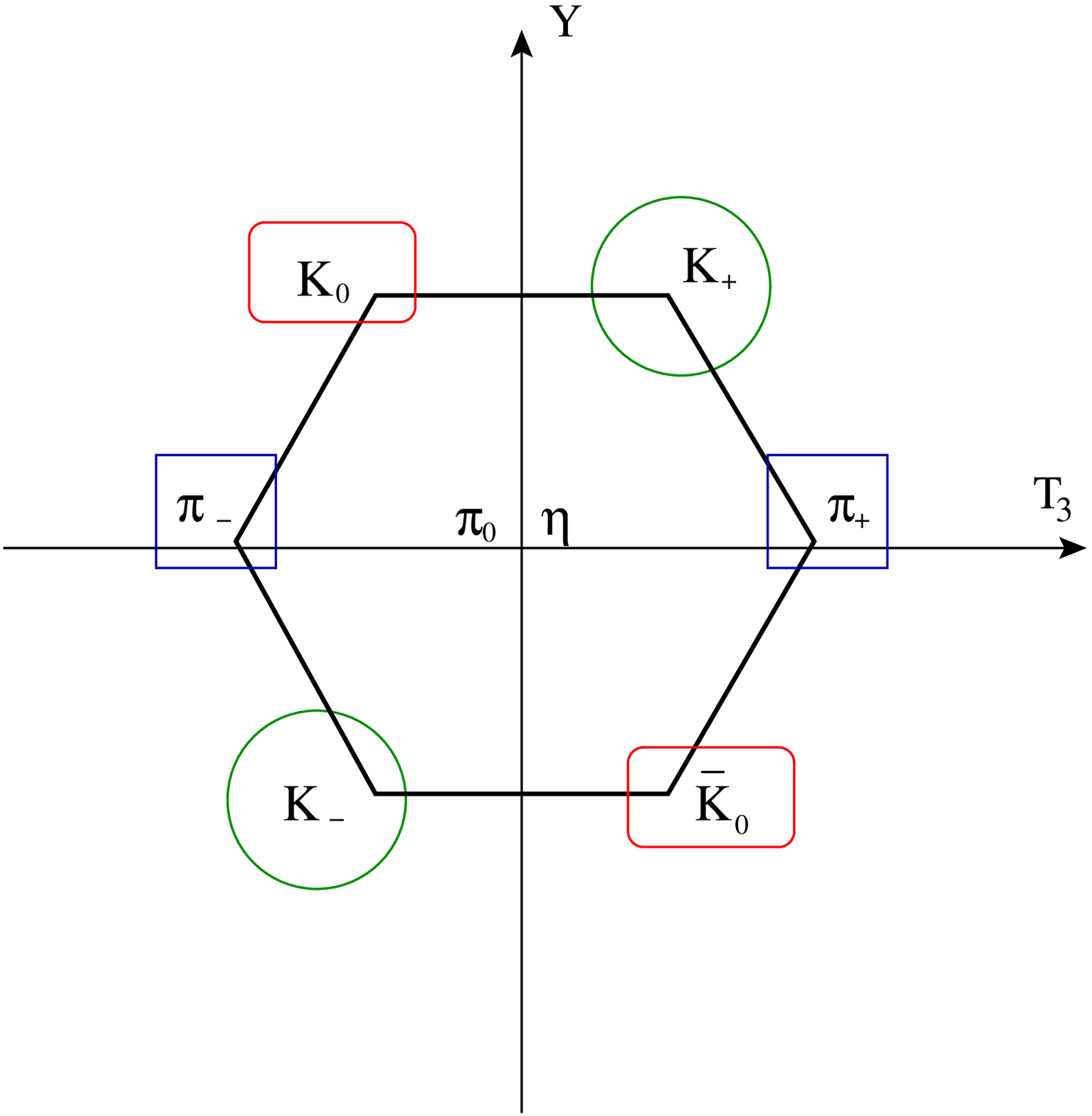}
\caption{(color online) Weight diagrams of the mesonic octet. In the top panel we have reported the axes corresponding to the third component of isospin (or $T$-spin), $U$-spin, $V$-spin and hypercharge, $Y$. Any mesonic state belongs to a multiplet of   $T$-spin, $U$-spin and $V$-spin as indicated in the diagram. In the bottom panel we have indicated the mesonic states having the same $T$-spin, $U$-spin and $V$-spin quantum numbers. Only  the states marked with the same symbol can mix.   The $\pi_0$ and the $\eta$  are not simultaneous    $T$-spin, $U$-spin and $V$-spin eigenstates; their mixing depends on the spontaneously induced charge of the vacuum. }
\label{fig:mixing}
\end{figure}

In the presence of background  isospin-rich matter or strangeness-rich matter, the Hamiltonian carries the third component of the isospin charge and of the strangeness charge (or hypercharge). The corresponding charges are explicitly broken, meaning that states with different third component of  isospin and different hypercharge can mix.  Indeed, the effect of a nonvanishing  $Q_Y$ and $Q_3$ is not only to produce a Zeeman-like splitting of the masses but also to tilt the vacuum in a certain direction in the isospin space corresponding to one of the nondiagonal generators of $SU(3)$. Let us discuss this issue more in detail. Given that the Hamiltonian has terms proportional to $T_3$ and $T_8$, the $SU(3)$ symmetry is explicitly broken. However, for  labeling   the mesonic  states we can use $T$-spin, $U$-spin and $V$-spin quantum numbers (actually only two of them are independent, indeed $T^2+U^2+V^2$ is one of the $SU(3)$ Casimir operators), because   $T^2$, $U^2$ and $V^2$ commute with $T_3$ and $T_8$. 
In Fig.~\ref{fig:mixing} we report the weight diagram of  the pseudoscalar mesonic octet. 
In the top panel  the axes correspond to  $T_3$, $U_3$, $V_3$ and $Y$ and the  values 
of the $T$-spin, $U$-spin and $V$-spin multiplets are reported. In the lower panel,  mesonic states that can mix  are marked with a different symbol.  These diagrams are valid both in the $\pi c$ phase and in the $K c$ phase. For example, from the top panel we see that charged pions can mix because they both have $T=1, U=1/2, V=1/2$, however charged pions cannot mix with kaons, because they all have $T=1/2$.  In Table~\ref{table:mixing}
we report the $T$ and $U$ quantum numbers of the  mesons with well defined $T$-spin and $U$-spin. In turn, the only allowed mixings are the followings:
$(\pi_+,\pi_-)$,  $(K_0,\bar K_0)$,  $(K_+,K_-)$, corresponding to  $(\phi_1, \phi_2)$,  $(\phi_4,\phi_5) $, $(\phi_6,\phi_7)$ mixing.  Regarding the  $\pi_0$ and the $\eta$, they have no well-defined $U$-spin or $V$-spin, thus a different reasoning must be used for understanding whether they mix or not. We will see that their mixing will depend on the particular spontaneously induced charge of the vacuum. 

\begin{table}[h!]
\begin{center}
\begin{tabular}{|c|c|}\hline
Mixing states & $(T,U)$\\ \hline
$\pi_+,\pi_-$ & $(1,1/2)$ \\ \hline
$K_+,K_-$ & $(1/2,1/2)$ \\ \hline
$K_0,\bar K_0$ & $(1/2,1)$ \\
\hline
%$\pi_0,\eta$ &  \\ \hline
\end{tabular}
\end{center}
\caption{Mixing mesons  with the corresponding $T$-spin and $U$-spin quantum numbers. These quantum numbers label the $SU(3)$ subspace spanned by the corresponding mesonic states.  The $\pi_0$ and the $\eta$ do not appear because they are not $U$-spin eigenstates.}
\label{table:mixing}
\end{table}% 

One of the important aspects is that the third component of isospin and the hypercharge  form the Cartan subalgebra of $SU(3)$, thus the associated charges cannot directy induce mixing between different states. In other words, the 
$Q_3$ and $Q_8$ charges can induce Zeeman-like mass splittings, but whether mixing between states will happen or not depends on the spontaneously induced  charge of the vacuum.   Note that  the operator associated to this induced charge  can be described in terms of lowering and raising operator of one of the  $SU(2) $ subgroups of $SU(3)$. 

Let us first consider the normal phase. In the normal phase there is no operator that can induce the mixing of the mesonic  states, thus the mesonic states remain unchanged but the $Q_3$ and $Q_8$ charges will induce Zeeman-like mass splittings.

In any of the condensed phases, there is an additional charge that is spontaneously induced, and the corresponding operator will lead to mixing.  %However, there will still be an $SU(2)$ subgroup that is left invariant. 

Let us first focus on isospin (or $T$-spin). We have to consider two cases.  Suppose that the vacuum has a charge that  commutes with $T^2$, as in the $\pi c$ phase,  say the charge corresponding to $T_2 = i( T_- - T_+)$, see Eq.~\eqref{pion}.  The $T_\pm$ operators can induce mixing among the charged pions and among the kaons. On the other hand, $T$-spin conservation does not allow the $\vert\pi_0\rangle =  \vert T=1, T_3=0\rangle$  to mix with the $\vert\eta\rangle = \vert T=0, T_3=0\rangle$.  

Suppose that the vacuum has a charge that  does not commute with $T^2$ as in the $Kc$ phase, see Eq.~\eqref{kaon}. Any operator that does not commute with isospin will commute with $U$-spin or with $V$-spin. In the $K c$ phase  $Q_5 \vert 0 \rangle \neq 0$, then the vacuum is not invariant under this charge. However, since  $[T_5,U]=0$  it follows that $U$-spin is conserved. The lowering and raising operator inducing the mixings will be $U_\pm$. Regarding the $\pi_0$ and the $\eta$, in this case  we have that $\vert U=1, U_3=0\rangle$ and $\vert U=0, U_3=0\rangle$ do not mix. Since $\vert U=1, U_3=0\rangle = \frac{\vert\pi_0\rangle + \sqrt{3}\vert \eta\rangle}{2}$ and $\vert U=0, U_3=0\rangle = \frac{\sqrt{3}\vert  \pi_0\rangle -\vert \eta\rangle}{2}$, these will be the mass eigenstates.   

In~\cite{Kogut:2001id} it was found a different mixing in both condensed phases, with just two blocks  $(\pi_+,\pi_-, \pi_0,\eta)$ and $(K_0,\bar K_0, K_+,K_-)$, meaning that states with different $T$-spin, $U$-spin and  $V$-spin mix. As we will see in the next section,  using the tree-level Lagrangian we find agreement with the mixing reported in Table~\ref{table:mixing}.

\subsection{Mesonic mass spectrum}
For vanishing chemical potentials and for equal light quark masses, the tree-level values of   the mesonic octet masses in $\chi$PT are known to be given by
\begin{align}
  m_\pi^2&= 2 G m/F_0^2\,,  \\
  m_K^2&= G (m+m_s)/F_0^2\,, \\
  m_{\eta^0}^2&= 2 G (m+2 m_s)/3 F_0^2=(4 m_K^2-m_\pi^2)/3\,,
\end{align}
where $m_s$ is the strange quark mass.
In the normal phase the effect of the isospin and strange quark chemical potential is a Zeeman-like mass splitting by contribution proportional to the isospin charge and strangeness,
\begin{align}
    m_{\pi^0}&=m_\pi\,, \\
    m_{\pi^\pm}&=m_\pi \mp \mu_I\,,\label{eq:pi+-masses}\\
    m_{\eta^0}&=    \sqrt{(4 m_K^2-m_\pi^2)/3}\,, \\
     m_{K^\pm}&=m_K\mp\frac12 \mu_I\mp\mu_S\,,  \\
    m_{K^0/\bar K^0}&=m_K\pm\frac12 \mu_I\mp\mu_S \,.
\label{eq:norm_masses}
\end{align} 

Below we will discuss the masses of the scalar mesons in the condensed phases.  To obtain the eigenstates we follow the same  procedure used in the two-flavor case, thus we first expand $\Sigma$ in order to obtain the quadratic terms in the fields then, if  necessary, we rescale and rotate  them to have canonical kinetic terms.  We will denote the mass eigenstates with a tilde. 

In Fig.~\ref{fig:masses} we report the obtained results for the  pseudoscalar mesonic octet masses  as a function of $\mu_I/m_\pi$ for three different values of the strange quark chemical potential. In the top panel we take $\mu_S=200$ MeV that is smaller than $\bar\mu_S =425$ MeV,  in the middle panel  we take  $\mu_S=460$ MeV and in the bottom panel $\mu_S=550$ MeV, which is the largest possible value of the strange quark chemical potential that can be considered in the present realization of $\chi$PT. The solid vertical lines correspond to the second order phase transitions between the normal phase and a condensed phase. The dashed vertical lines correspond to  first order phase transitions between the $K c$ phase and the $\pi c$ phase.  The top panel and the middle panel should be compared with the corresponding results reported in~\cite{Kogut:2001id} in Fig.~4 and in Fig.~5, respectively. In the normal phase our results agree with those of~\cite{Kogut:2001id}, but in both  condensed phases we disagree with  the results reported  in~\cite{Kogut:2001id}.  As already discussed, this is due to the fact that we find a different mixing pattern, even if we use the same model of $\chi$PT of~\cite{Kogut:2001id}. For this reason we  discuss our results in detail.

\subsubsection{Pion condensation phase}
We find mixing within the following pairs of states: $(\phi_1, \phi_2),~(\phi_4,\phi_5) $ and $(\phi_6,\phi_7)$, while the $\phi_3$ and $\phi_8$ fields do not mix.   Thus, in agreement with the discussion in Sec.~\ref{sec:mixing}, we do not find mixing between the $\vert \pi_0 \rangle$ and the $\vert \eta \rangle$. The reason is that in the $\pi c$ phase $T$-spin is conserved and therefore  the  $\vert \pi_0 \rangle = \vert T=1, T_3=0 \rangle$ state  and the $\vert \eta \rangle = \vert T=1, T_3=0 \rangle$ state cannot mix. 
 
Since the Lagrangian can be organized in a block diagonal form, we can treat separately the various sectors. 
By the field rescaling
\begin{align}
\phi_{1,3}&\rightarrow \tilde\phi_{1,3}=\phi_{1,3} \cos{\alpha_\pi}\,, \\
\phi_{4,5,6,7}&\rightarrow\tilde\phi_{4,5,6,7}=\phi_{4,5,6,7} \cos \(\frac{\alpha_\pi}{2}\)\,,
\label{eq:rescalate1}
\end{align} we obtain canonical kinetic terms.
As in Sec. \ref{quadL} it is useful to turn to momentum space, so that one can absorb the  terms linear in energy in the   propagator. In this way we obtain the canonical Lagrangian
\be
{\cal L}= \tilde\Phi^t\text{diag} \left(S^{-1}_{12},  S^{-1}_{45}, S^{-1}_{67}, S^{-1}_{3}, S^{-1}_{8} \right) \tilde\Phi\,,
\ee
where $ \tilde\Phi = (\tilde\phi_1, \tilde\phi_2, \tilde\phi_4, \tilde\phi_5, \tilde\phi_6, \tilde\phi_7, \tilde\phi_3, \tilde\phi_8)$. 

Regarding the  $(\tilde\phi_1, \tilde\phi_2)$ sector, the mixing is the same obtained in the two-flavor case, thus the results obtained in Sec.~\ref{sec:twoflavor} hold unchanged. The same applies to the $\tilde\phi_3$ sector. For the $\tilde\phi_8$ sector corresponding to the $\eta$ field we obtain
\be
m_{\tilde{\eta}}^2=m_{\eta}^2+\frac{1}{3} m_{\pi}^2\(\frac{m_\pi^2-\mu_I^2}{\mu_I^2}\)\,.
\ee
This expression is rather different from the corresponding expression given in Eq.~(27) of~\cite{Kogut:2001id}. As already mentioned   in the $\pi c$ phase we do not find tree-level mixing between the $\eta$ and any other meson, while in~\cite{Kogut:2001id} the $\eta$ mixes with the pions.

For the $(\tilde\phi_4,\tilde\phi_5)$ sector  we have
\be
S^{-1}_{45}=\left( \begin{array}{cc}
                            k^2-M_4^2 & -i k_0 s_{45} \\
                             i k_0 s_{45} & k^2-M_5^2 
                            \end{array} \right)\,,
\ee
where
\begin{align}
s_{45}&=\mu_I\cos\alpha_\pi +2\mu_S\,,  \\
M_4^2&=M_5^2=m_k^2+\frac{1}{4}\mu_I^2-\mu_S^2-\frac{1}{2}(\mu_I^2+2\mu_I\mu_S)\cos\alpha_\pi\,.
\end{align}
The masses of the rotated kaons are given by
\begin{align}
m_{\tilde{K}^-/\tilde{K}^+}&=\pm\frac{1}{2}\(\frac{m_\pi^2}{\mu_I}+2\mu_S\pm\sqrt{\(\frac{m_\pi^2}{\mu_I}+2\mu_S  \)^2+4 M_4^2}  \) 
\end{align}
For the $(\tilde\phi_6,\tilde\phi_7)$, we have
\be
S^{-1}_{67}=\left( \begin{array}{cc}
                            k^2-M_6^2 & -i k_0 s_{67} \\
                             i k_0 s_{67} & k^2-M_7^2 
                            \end{array} \right)\,,
\ee
where
\begin{align}
s_{67}&=\mu_I \cos \alpha_\pi-2\mu_S,  \\
M_6^2&=M_7^2=m_k^2+\frac{\mu_I^2}{4}-\mu_S^2-\frac{\mu_I^2-2\mu_I\mu_S}{2}\cos \a_\pi\,.
\end{align}
The masses of the rotated neutral kaons are given by
\begin{align}
m_{\tilde{K}^0/\tilde{\bar{K}}^0}&=\pm\frac{1}{2}\(\frac{m_\pi^2}{\mu_I}-2\mu_S\pm\sqrt{\(\frac{m_\pi^2}{\mu_I}-2\mu_S  \)^2+4 M_6^2}  \)\,. 
\end{align}
\subsubsection{Kaon condensation phase}
In agreement with the discussion in Sec.~\ref{sec:mixing},  in the kaon condensation phase we find mixing between $(\phi_1, \phi_2)$, $(\phi_3,\phi_8)$, $(\phi_4,\phi_5)$ and $(\phi_6,\phi_7)$. By the field transformation  
\begin{align}
\phi_{1,2,6,7}&\rightarrow\tilde\phi_{4,5,6,7}=\phi_{4,5,6,7} \cos \(\frac{\alpha_K}{2}\)\,, \nonumber \\
\phi_{4}&\rightarrow \tilde\phi_{4}=\phi_{4} \cos{\alpha_K}\,,  \\
\left( \begin{array}{c} \tilde\phi_3 \\ \tilde\phi_8 \end{array} \right)&=\left( \begin{array}{cc} -\frac{\sqrt{3}}{2} & \frac{1}{2} \\ \frac{1}{2}\cos \a_K & \frac{\sqrt{3}}{2}\cos \alpha_K \end{array} \right) \left( \begin{array}{c} \phi_3 \\ \phi_8 \end{array} \right)\,, \nonumber
\end{align}
we obtain a  block diagonal Lagrangian in momentum space 
\be
{\cal L}= \tilde\Phi\text{diag} \left(S^{-1}_{12}, S^{-1}_{45}, S^{-1}_{67}, S^{-1}_{38} \right) \tilde\Phi^t\,.
\ee
As anticipated in Sec.~\ref{sec:mixing} the $\pi_0$ and the $\eta$ mix, and according to that discussion  the mixed states are proportional  to the $U$-spin eigenstates, indeed $\vert\tilde\phi_3\rangle = \vert U=0, U_3=0\rangle$ and   $\vert\tilde\phi_8\rangle =\cos\alpha_K \vert U=1, U_3=0\rangle$.

For the  $(\tilde\phi_1,\tilde\phi_2)$ sector we obtain
\be
S^{-1}_{12}=\left( \begin{array}{cc}
                            k^2-M_1^{'2} & -i k_0 u_{12} \\
                             i k_0 u_{12} & k^2-M_2^{'2} 
                            \end{array} \right)\,,
\ee
where
\begin{align}
u_{12}&=\frac{1}{2}((3+\cos \a_K)\mu_I+2(\cos \a_K-1)\mu_S)\,,   \\
M_1^{'2}&=M_2^{'2}=m_\pi^2-\frac{1}{2}\mu_I^2(1+\cos \a_K)-4\mu_I\mu_S(\cos \a_K-1)\,,    \nonumber
\end{align}
and we find the pion masses
\begin{widetext}
\begin{align}
m_{\tilde\pi^\pm}&= 
\mp\frac{1}{2}\(\frac{\mu_I}{2}(3+\cos \a_K)+\mu_S(\cos \a_K-1)\mp\sqrt{\(\frac{\mu_I}{2}(3+\cos \a_K)+\mu_S(\cos \a_K-1)  \)^2+4 M_1^{'2}} \)\,. 
\end{align}
\end{widetext}

For the  $(\tilde\phi_4,\tilde\phi_5)$ sector  we obtain
\be
S^{-1}_{45}=\left( \begin{array}{cc}
                            k^2 & -i k_0 u_{45} \\
                            i k_0 u_{45} & k^2-M_5^{'2} 
                            \end{array} \right)\,,
\ee
where
\begin{align}
M_5^{'2}&= m_k^2 \cos \a_K-\frac{1}{4}(\mu_I+2\mu_S)^2  \cos (2\a_K)\,,  \\
u_{45}&= \frac{1}{4} \( \frac{1+3\cos \a_K)}{\cos \a_K}\) (\mu_I+2\mu_S)\,,  
\end{align}
and the corresponding  masses are given by
\begin{align}
m_{\tilde{K}^+}&=0\,,    \\
m_{\tilde{K}^-}&= \sqrt{M_5^{'2}+u_{45}^2}\,.    
\end{align}

For the $(\tilde\phi_6,\tilde\phi_7)$ sector we obtain
\be
S^{-1}_{67}=\left( \begin{array}{cc}
                            k^2-M_6^{'2} & -i k_0 u_{67} \\
                            i k_0u_{67} & k^2-M_7^{'2} 
                            \end{array} \right)\,,
\ee
where
\begin{align}
M_6^{'2}&=M_7^{'2}= m_k^2+\frac{\mu_I-2\mu_S}{4}((\cos\a_K (\mu_I+2\mu_S)-2\mu_I)\,,   \\
u_{67}&=\frac{1}{2}((-3+\cos\a_K)\mu_I+2(1+\cos\a_K)\mu_S)\,,
\end{align}
and we find the masses of the neutral kaons
\begin{align}
m_{\tilde{\bar{K}}^0/\tilde K^0}&=\pm\frac{1}{2} (u_{67}\pm\sqrt{u_{67}^2+4M_6^{'2}})\,.%\\ 
%m_{\tilde K^0} &=-\frac{1}{2}  (u_{67}-\sqrt{u_{67}^2+4M_6^{'2}})\,.  
\end{align}

For the  $(\tilde\phi_3,\tilde\phi_8)$ sector we have
\be
S^{-1}_{38}=\left( \begin{array}{cc}
                            k^2-M_3^{'2} &  u_{38} \\
                            u_{38} & k^2-M_8^{'2} 
                            \end{array} \right)\,,
\ee
where
\begin{align}
M_3^{'2}&=\frac{1}{24}\left[ \cos \alpha_K\left(2  m_k^2 +9  m_\pi^2+6 \frac{G}{F_0^2}m_s\right)\right.\nonumber\\ &\left.+m_\pi^2 (16 -6   \cos^2 \a_K)\right] \,,  \\
M_8^{'2}&=\frac{1}{8 \cos \a_K}(2 m_k^2+(1+2 \cos\a_K)m_\pi^2\nonumber\\ &+6\frac{G}{F_0^2}m_s \cos \a_K)\,, \\
u_{38}&=\frac{1}{8 \sqrt{3}}(-2m_k^2+(3+2\cos\a_K)m_\pi^2-6 \frac{G}{F_0^2} m_s)\,, 
\end{align}
and we find the masses
\begin{align}
m_{\tilde\pi^0/\tilde\eta}&=\sqrt{\frac{M_3^{'2}+M_8^{'2}\mp\sqrt{4 u_{38}^2+(M_3^{'2}-M_8^{'2})^2}}{2}}\,.   %   \\
%m_{\tilde\eta}&= \sqrt{\frac{M_3^{'2}+M_8^{'2}+\sqrt{4 u_{38}^2+(M_3^{'2}-M_8^{'2})^2}}{2}}\,.     \nonumber
\end{align}

\begin{figure}[th!]
\includegraphics[width=8.cm]{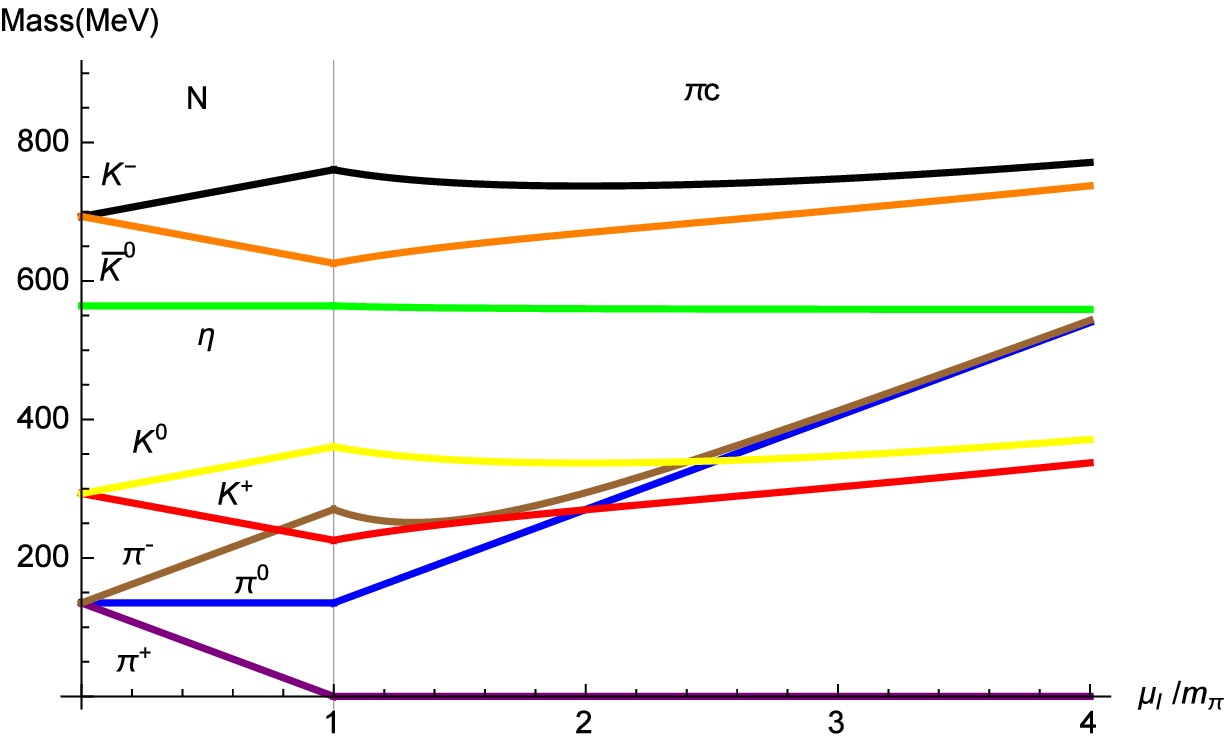}
\includegraphics[width=8.cm]{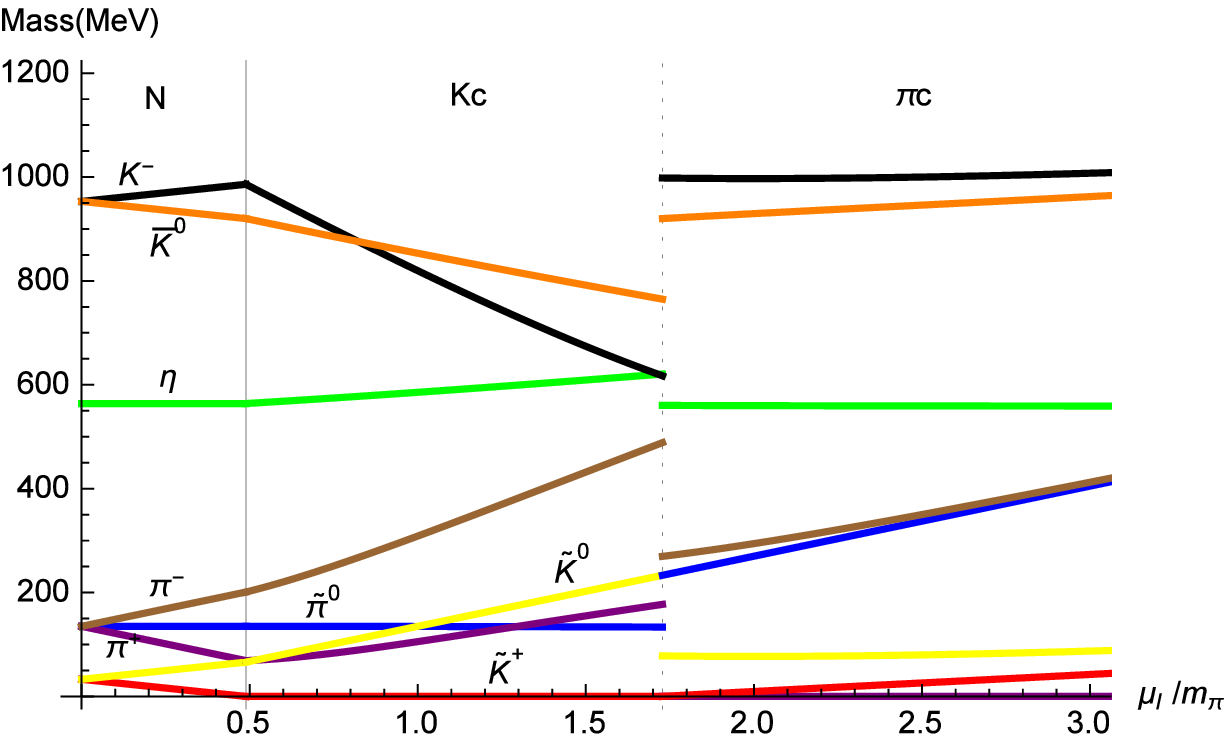}
\includegraphics[width=8.cm]{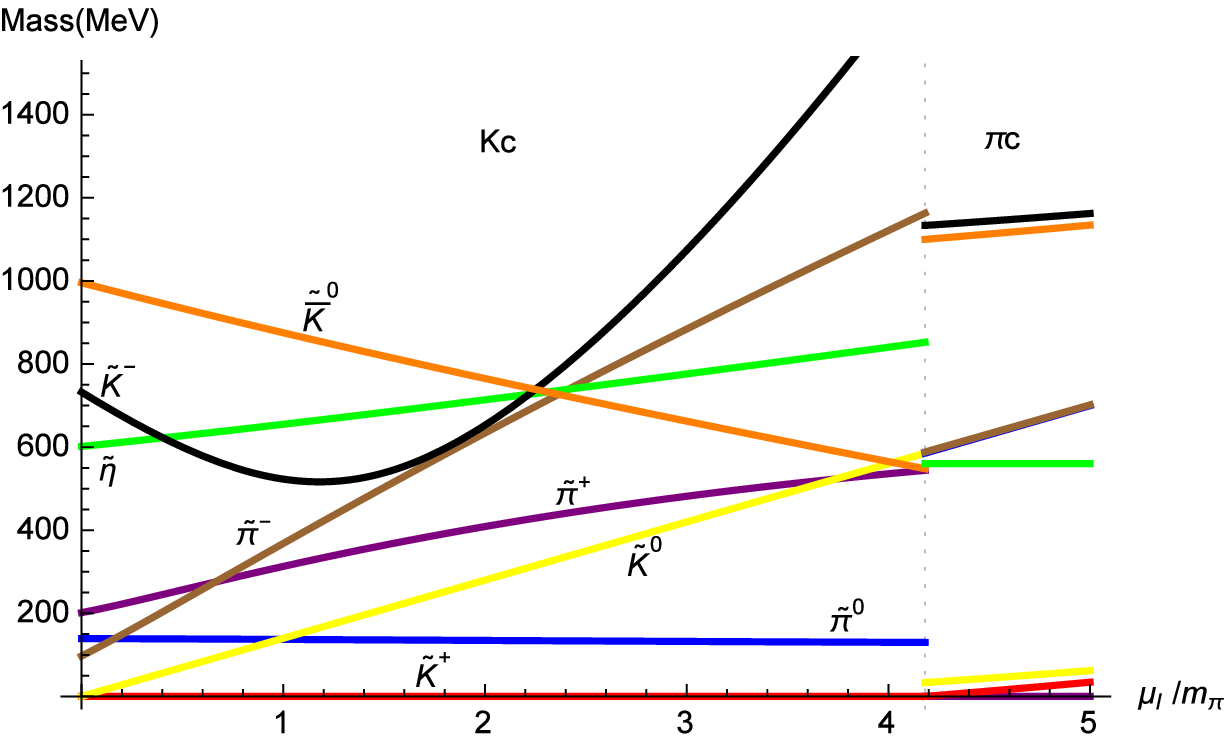}
\caption{(color online) Mass spectrum of the pseudoscalar mesonic octet.   Top panel, results obtained  for $\mu_S=200$ MeV. The vertical solid line represents the second order phase transition between the normal phase and the pion condensation phase.  In this case the strange quark chemical potential is below the threshold value for kaon condensation, $425$ MeV, thus the kaon condensed phase does not take place for any value of $\mu_I$.  Middle panel, results obtained  for  $\mu_S=460$ MeV. The vertical solid line represents the second order phase transition from the normal phase to the kaon condensation phase. The vertical dotted line corresponds to the first order phase transition between the kaon condensed phase and the pion condensed phase. Bottom panel, results obtained  for  $\mu_S=550$ MeV, corresponding to the largest value of $\mu_S$. Also in this case the vertical dotted line corresponds to the first order phase transition between the kaon condensed phase and the pion condensed phase. }
\label{fig:masses}
\end{figure}

Several remarks are in order: 
\begin{enumerate}
\item The masses are continuous across the second order phase transition  but may have a jump at the first order phase transition line. Actually, at the first order phase transition point all the masses are discontinuous but the $\tilde K^+$
mass, which is the pseudo NGB  associated to the superfluid mode of the $Kc$ phase.
\item The mass hierarchy can change dramatically with some kaons becoming lighter than pions. This has the effect of forbidding certain decaying process and/or allow decay process that are not allowed for vanishing chemical potentials. We will discuss some of these processes in Sec.~\ref{sec:piondecays}.
\item The light charged states can become absolutely  stable.
\item Some mesonic masses in Fig.~\ref{fig:masses}  are about the $\chi$PT breaking scale $\sim 1$ GeV. In this range of energies the theory is not under quantitative control, however we expect that the obtained mass hierarchy remains qualitatively the same.
\end{enumerate}

\section{Mesonic decays}\label{sec:piondecays}
In this section we discuss the  decay rate of the  pions in the normal phase and in the $\pi c$ phase. For definiteness we consider $\mu_S=0$, however the same results hold for any $\mu_S < \bar\mu_{S}$ .
It is important to stress that some processes are sensitive to the fact that matter is in electroweak equilibrium.  
The weak decay processes
\begin{align}
u &\rightarrow d + \ell^+ +\nu_\ell \,,\\
d + \ell^+ &\rightarrow u + \bar\nu_\ell \,,
\end{align}
where $\ell^\pm$ indicates a charged leptonic species and $\nu_\ell$ the corresponding neutrino, impose that $\mu_\ell^+ = \mu_I$ (assuming that neutrinos are not trapped). Therefore, for $\mu_I>0$ the leptonic decay of  positively charged mesons can be Pauli blocked. However, if charged leptons are not trapped (as in heavy ions), then  $\mu_\ell^+ =0$. In the following we will consider both cases and compare the results obtained for $\mu_\ell^+ =0$ with those obtained for   $\mu_\ell^+ = \mu_I$.

\subsection{Leptonic decays}
We first consider the leptonic decays, which are the dominant decay channels in vacuum. We separately consider the normal phase and the $\pi c$ phase.

 \subsubsection{Normal phase}
\begin{figure}[t]
%\centering
\includegraphics[width=8.cm]{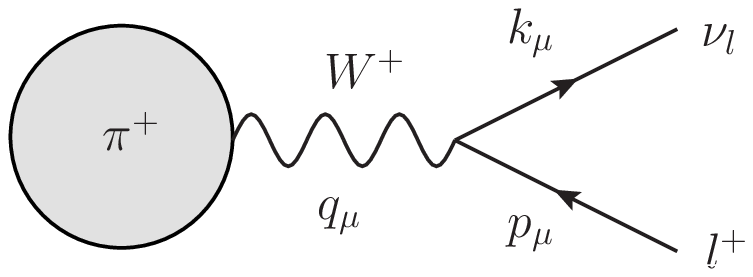}
\includegraphics[width=8.cm]{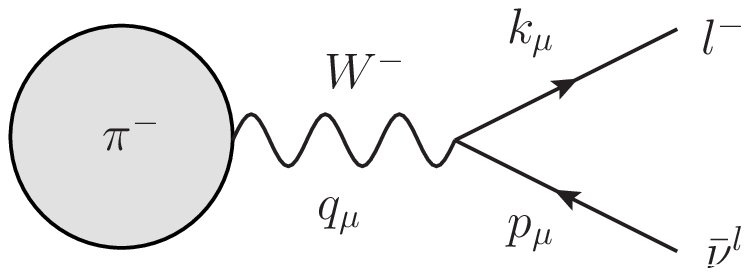}
\caption{Feynman diagrams representing the charged pion decays in the normal phase.}
\label{fig:feyn1}
\end{figure}

In the normal phase the meson mass eigenstates are also the charge eigenstates, thus the standard leptonic decay channels
\begin{align}
\pi^+ &\rightarrow \ell^+ \nu_\ell\,,  \\
\pi^- &\rightarrow \ell^- \bar{\nu}_\ell\,,
\end{align}
are relevant and the corresponding  diagrams are reported in Fig.~\ref{fig:feyn1}.  
For $\mu_{\ell^+} =0  $ we obtain 
\begin{align}
\frac{\Gamma_{\pi^+\rightarrow \ell^+\n_\ell}}{\Gamma^0_{\pi \rightarrow \ell \nu_\ell}} &=\frac{m_{\pi^+}}{m_\pi} \left(\frac{1-m_\ell^2/m_{\pi^+}^2}{1-m_\ell^2/m_{\pi}^2}\right)^2\,,\\
\frac{\Gamma_{\pi^-\rightarrow \ell^-\bar\n_\ell}}{\Gamma^0_{\pi \rightarrow \ell \nu_\ell}} &=\frac{m_{\pi^-}}{m_\pi} \left(\frac{1-m_\ell^2/m_{\pi^-}^2}{1-m_\ell^2/m_{\pi}^2}\right)^2\,,
\end{align}
where  $m_{\pi^\pm}$  are the masses of the charged pions in the normal phase, given in Eq.~\eqref{eq:pi+-masses}, and 
\be\label{eq:gamma0}
\Gamma^0_{\pi \rightarrow \ell \nu_\ell}=\frac{G_F^2 F_0^2 V_{ud}^2 m_\ell^2 m_{\pi}}{4 \pi} \( 1-\frac{m_\ell^2}{m_{\pi}^2} \)^2\,,
\ee 
is the standard leptonic decay width, with  $G_F$  the Fermi constant and $V_{ud}$  the $ud$ element of the CKM matrix.

The behavior of the leptonic decay channels is shown in Fig.~\ref{fig:gammas}, the normal phase corresponds to $\mu_I/m_\pi <1$. In the top panel are reported the results obtained for $\mu_{\ell^+}=0$; in the bottom panel the results obtained for $\mu_{\ell^+}=\mu_I$. Note that the decay width in the normal phase is not affected by the ${\ell^+}$ chemical potential. With increasing $\mu_I$ the mass of the  $\pi^+$  decreases and its 
decay width vanishes at the point in which the available phase space shrinks to zero, corresponding to
\be
\frac{\mu_I}{m_\pi} = 1 - \frac{m_\ell}{m_\pi}\,.
\ee
Thus the decay width  $\pi^+\to \mu^+ \nu_\mu$ vanishes at  $\mu_I/m_\pi \simeq 0.245$. Close to this value, for  $\mu_I/m_\pi \gtrsim 0.241$, the decay $\pi^+\to e^+ \nu_e$ becomes the dominant process. The positron decay channel closes only  for $\mu_I/m_\pi \simeq 0.996$. Note that with increasing $\mu_I$ the mass of the $\pi^-$  increases, thus the width of the  leptonic decay $\pi^-\to \mu^- \bar\nu_\mu$ increases. 
\subsubsection{Pion condensation phase}
In the $\pi c$ phase the $\tilde\pi^+$ field is massless, thus it will not decay. On the other hand $\tilde\pi^-$ is a combination of $\pi^+$ and $\pi^-$ and it can decay in both $\ell^+\n_\ell$ and $\ell^-\bar{\n}_\ell$. We can describe this decay process as given by the linear combination of the two charge eigenstates as in Fig.~\ref{fig:cond_decay}.

\begin{figure}[t]
\centering
\includegraphics[width=8.cm]{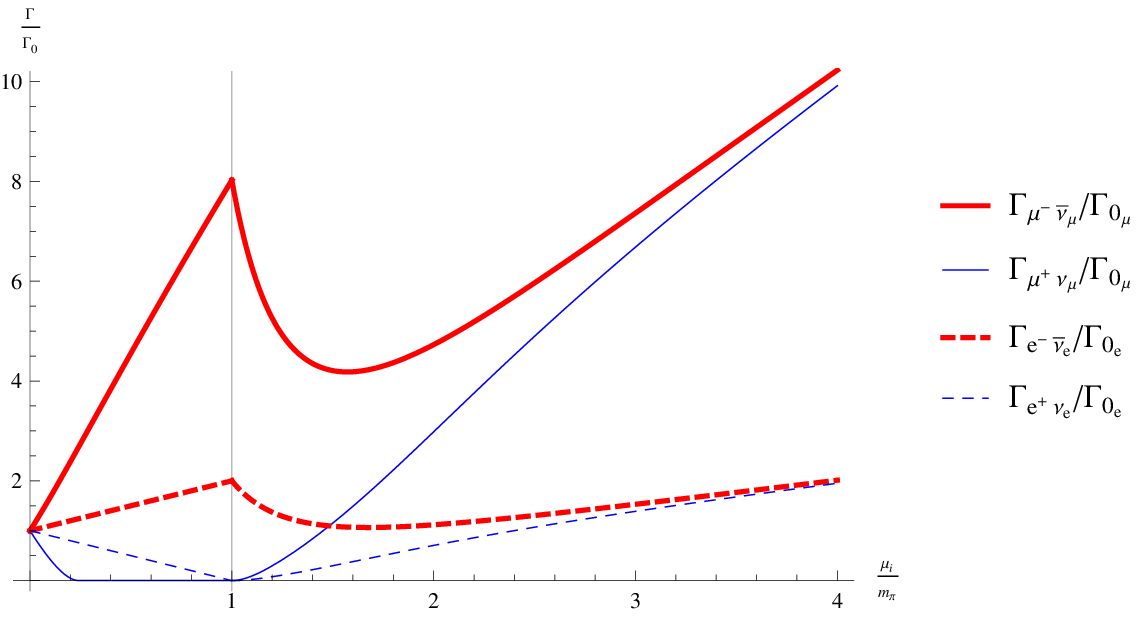}
\includegraphics[width=8.cm]{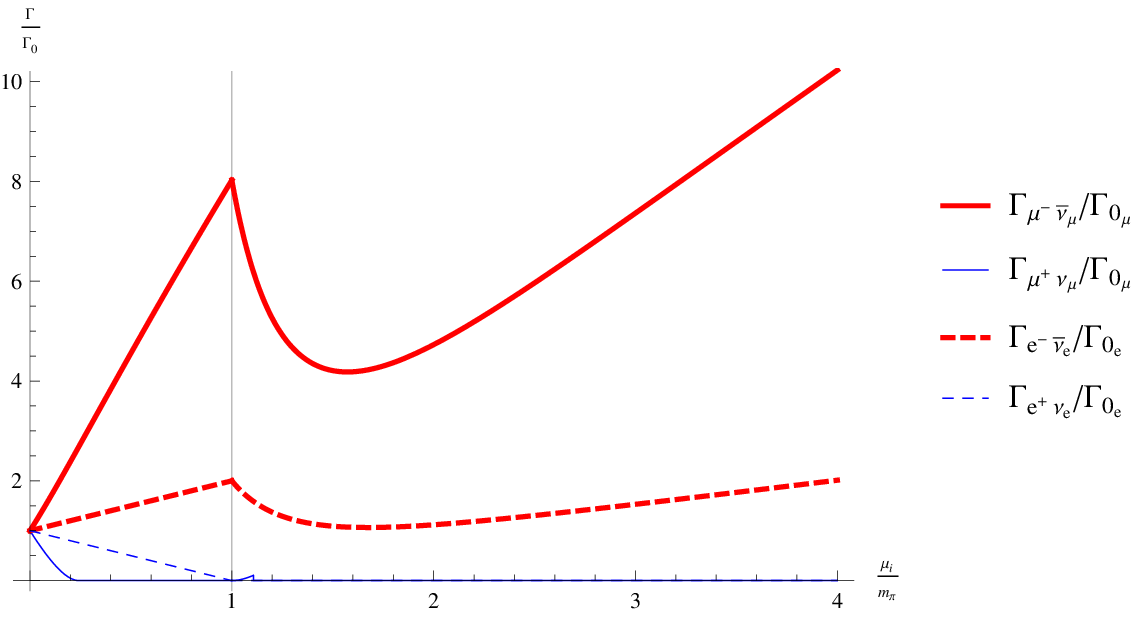}
\caption{(color online) Normalized leptonic decay rates of charged pions in normal phase  and in the  condensed phase normalized to the value in vacuum. The phase transition between the normal phase and the pion condesed phase corresponds to the  solid vertical line. Top, results obtained for vanishing leptonic chemical potential. Bottom, results obtained assuming weak equilibriub. In this case the decay in  positively charged leptons is Pauli blocked. In both plots the thick solid line represents $\Gamma_{\m^-\bar{\n}_\m}/\Gamma_{0_\m}$, the thin solid line represents  $\Gamma_{\m^+\n_\m}/\Gamma_{0_\m}$, the thick dashed line represents  $\Gamma_{e^-\bar{\n}_e}/\Gamma_{0_e}$ and the thin dashed line represents  $\Gamma_{e^+\n_e}/\Gamma_{0_e}$}
\label{fig:gammas}
\end{figure}

  Using Eq.~\eqref{U} to describe this linear combination, we obtain the vertex factors
\begin{align}
\hat\Gamma^\mu_{\tilde{\pi}^-W_\m^+}&=-\frac{F_0 g V_{ud}}{2 \sqrt{2}}(-i q^\m)(U_{21}^* \cos \a_\pi+i U^*_{22})\,, \label{eq:vert1}\\
\hat\Gamma^\mu_{\tilde{\pi}^-W_\m^-}&=-\frac{F_0 g V_{ud}}{2 \sqrt{2}}(-i q^\m)(U_{21}^* \cos \a_\pi-i U^*_{22})\,,
\label{eq:vert2}
\end{align}
where $g$ is the weak coupling constant. More details on the derivation of Eq.~\eqref{eq:vert1} and~\eqref{eq:vert2}  are given in the Appendix \ref{appA}. The tensorial structure of these vertices is unchanged from the normal phase, see for example~\cite{Scherer:2002tk}. The only difference is in the appearance of the $U$-matrix elements, corresponding to the mixing of the fields,  and in the $\cos \a_\pi$ coefficients,  arising from   the normalization of the fields. Thus, the tensorial contribution to $|\mathcal{M}^2|$ will be unchanged from the normal phase and we only have to take into count the different mixing coefficients. Before giving the results for the decay rates in the condensed phase, we want to remark that the $U$ matrix in Eq.~\eqref{U} is $q_0$ dependent: the mixing depends on the energy of the $W$ boson (that is the energy of the decaying pion). This energy dependence does not complicate our calculation because when we integrate the squared amplitude to obtain the decay width, we have to choose a frame to express the kinematical variables: we have chosen the charged pion rest frame,  meaning that $q_\m=(m_{\tilde{\pi}_-},{\bm 0})$. For this reason, $U_{21}^*$ and $U_{22}^*$ will depend only on $\mu_I$. In turn, the leptonic decay rates are given by
\begin{align}
\frac{\Gamma_{\tilde{\pi}_-\rightarrow \ell^+\nu_\ell}}{\Gamma^0_{\pi \rightarrow \ell \nu_\ell}}&=\frac{|U_{21}^* \cos \a+i U^*_{22}|^2}{2}\frac{m_{\tilde{\pi}^-}}{m_\pi} \left(\frac{1-m_\ell^2/m_{\tilde\pi^-}^2}{1-m_\ell^2/m_{\pi}^2}\right)^2\,,\\
\frac{\Gamma_{\tilde{\pi}_-\rightarrow \ell^-\bar\nu_\ell}}{\Gamma^0_{\pi \rightarrow \ell \nu_\ell}}&=\frac{|U_{21}^* \cos \a-i U^*_{22}|^2}{2}\frac{m_{\tilde{\pi}^-}}{m_\pi} \left(\frac{1-m_\ell^2/m_{\tilde\pi^-}^2}{1-m_\ell^2/m_{\pi}^2}\right)^2\,,
\end{align}
where $\Gamma^0_{\pi \rightarrow \ell \nu_\ell}$ is given in Eq.\eqref{eq:gamma0}.

The decay width of both channels are reported in Fig.~\ref{fig:gammas}; the $\pi c$ phase corresponds to $\mu_I/m_\pi >1$. The Pauli blocking suppresses the $\ell^+ \nu_\ell$ channel but does not affect the $\ell^- \bar \nu_\ell$ channel. Regarding the latter, it is curious to note the presence of a local minimum around $\mu_I \sim 1.5 m_\pi$, which corresponds to the local minimum of the  $\tilde\pi^-$ mass located at $\mu_I/m_\pi = 3^{1/4}$ see Eq.~\eqref{eq:Etildepimeno}. The small difference between these two values is due to the mixing angle between the $\tilde\pi_-$ state and the  $\pi_-$ state and to the fact that the decay width is not a linear function of $\mu_I/m_\pi$.

\begin{figure}[h!]
%\centering
\includegraphics[width=8.cm]{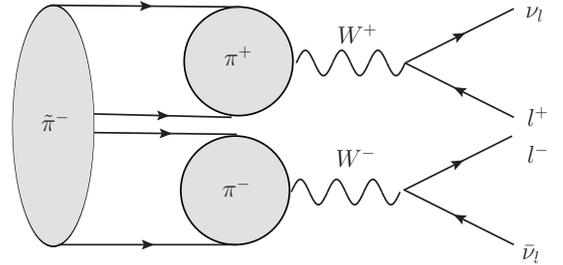}
\caption{Feynman diagram describing the $\tilde\pi^-$ decay channels as superposition of the decays of the charged pion eigenstates. The decays of the charged pions are as in Fig.~\ref{fig:feyn1}.   }
\label{fig:cond_decay}
\end{figure} 

\subsection{Semileptonic decays}
Regarding the semileptonic decays, one has to consider the  mass splitting between the charged pions and the  neutral pion due to unequal light quark masses, $\delta m_\pi = \left(m_{\pi_+}-m_{\pi_0}\right)\vert_{\mu_I=0}\sim 4.5$ MeV, which allows the  $\pi_{e3}$ decays  $\pi_+ \to \pi_0 e^+ \nu$ and $\pi_- \to \pi_0 e^- \bar\nu$ in vacuum. It  is interesting to note that  the former is kinematically forbidden for  $\Delta m_{\pi_+} = \left(m_{\pi_+}-m_{\pi_0}\right)\vert_{\mu_I} < m_e$ corresponding to  $\mu_I \gtrsim \delta m_\pi  - m_e \simeq 4$ MeV.  Considering the Pauli blocking effect on   positrons, the process is forbidden for $\Delta m_{\pi^+}< \mu_I$, corresponding to $\mu_I \gtrsim \delta m_\pi/2 \simeq 2.3$ MeV. On the other hand, the  $\pi_- \to \pi_0 e^- \bar\nu$ is enhanced by the isospin chemical potential and no Pauli blocking effect is present. Neglecting the recoil of the $\pi_0$ and considering the leading order in  $\Delta m_{\pi^-}/ m_\pi$ and  $m_e/\Delta m_{\pi^-}$, we obtain
\be
\frac{\Gamma_{\pi_-\to \pi_0 + e +\bar\nu_e}}{\Gamma^0_{\pi_-\to \pi_0 + e +\bar\nu_e}} = \left( \frac{\delta m_\pi + \mu_I}{\delta m_\pi}\right)^5\,,
\ee
meaning that the corresponding decay width is largely enhanced with respect to the decay in vacuum. However, it turns out that this $\pi_{e3}$ channel is still suppressed by at least two orders of magnitude with respect to the corresponding leptonic decay. 

\begin{figure}[t]
\centering
\includegraphics[width=8.cm]{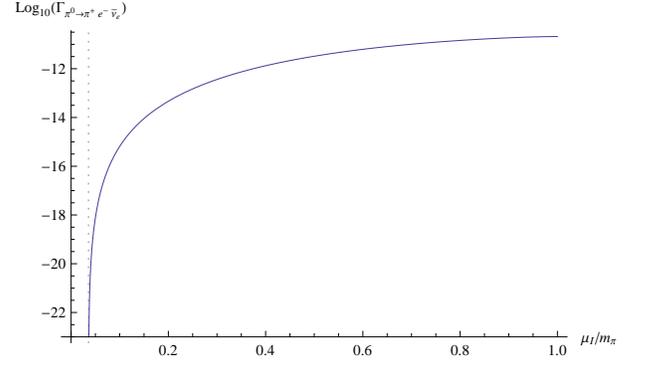}
\caption{Semileptonic  decay rates in normal  phase for the process~\eqref{eq:semilept-pi0}. The vertical axes units are MeV, thus the width changes between $\sim 10^{-22}$ MeV and $\sim 10^{-11}$ MeV. This decay channels becomes kinematically allowed for $\mu_I \gtrsim 5$ MeV, corresponding to the vertical dotted line.  }
\label{fig:gammasemileptonic}
\end{figure}

More interesting is perhaps the fact that the processes 
\be \pi^0 \to \pi^+ + e^-+ \bar\nu_e\label{eq:semilept-pi0}\,,\ee 
and 
\be\pi^0 \to \pi^+ + \mu^-+ \bar\nu_\mu\,,\ee are kinematically allowed for $\mu_I >\delta m_\pi + m_e \simeq 5$ MeV and  $\mu_I >\delta m_\pi + m_\mu \simeq 110$ MeV respectively. These semileptonic decays are forbidden in  vacuum, but are allowed in isospin rich matter.  In Fig.~\ref{fig:gammasemileptonic} we report the semileptonic decay rate for the process \eqref{eq:semilept-pi0} in the normal phase. Note that the decay width for $\mu_I/m_\pi>0.5$ leads to a mean lifetime $1/\Gamma \sim 10^{-10}$ s.
In the $\pi c$ phase  from dimensional analysis (neglecting the electron mass) we have that 
\be \Gamma_{\pi^0 \to \pi^+ + e^-+ \bar\nu_e} \propto G_F^2 \mu_I^5\sim 10^{-11} \left(\frac{\mu_I}{m_\pi}\right)^5 \text{MeV}\,,\label{eq:gammapi0semilept}\ee
therefore this decay width is comparable with that of the normal phase for $\mu_I/m_\pi>0.5$. From the above reasoning it is clear that
  the chain of semileptonic decays 
\begin{align}
\pi^- &\to \pi^0 + e + \bar\nu_e\,, \\
\pi^0 &\to  \pi^+ + e +\bar\nu_e\,,
\end{align}
 feed the $\pi_+$ states, which are stable.

\subsubsection{Neutral pion decay}
Let us briefly comment on the 
\be
\pi_0 \to \gamma \gamma \,,
\ee
decay. In the normal phase this process is the dominant decay channel for neutral pions. Since in the broken phases  the photon has a screening mass, see Eq.~\eqref{eq:screening}, one might naively expect that this channel is suppressed. However, in the $\pi c$ phase $m_{\pi_0}= \mu_I$, see~Eq.~\eqref{eq:pi0}, thus  this process is kinematically allowed if
\be
\mu_I > 2\, F_0\, e \sin \a_\pi\,,
\ee
that is always satisfied. Therefore, we expect that this decay channel has a width of the order of the width in vaccum, meaning that it should be  larger than the semileptonic width in Eq.~\eqref{eq:gammapi0semilept}.

\section{Conclusions}\label{sec:conclusion}
We have studied various aspects of pseudoscalar mesons as a function of the isospin chemical potential and of the strange quark chemical potential. We have determined the mass eigenstates, obtaining results in disagreement with those of~\cite{Kogut:2001id}.  In particular in Sec.~\ref{sec:mixing} we have found the mixing pattern reported in Table~\ref{table:mixing} and we have determined under which circumstances the $\pi_0$-$\eta$ mixing happens. Since we have  obtained these results by theory group methods, they are expected to  hold in any theory describing mesonic states in the pion condensed phase and in the kaon condensed phase.  We have substantiated these results considering the low-energy chiral Lagrangian describing the interaction of mesonic states with an isospin and strangeness charged background. Note that this is the same low-energy Lagrangian used in~\cite{Kogut:2001id}.  

We have analyzed several pion decay channels finding a nontrivial behavior across the second order phase transition between the normal phase and the pion condensed phase. The semileptonic decays become efficient in populating the stable charged mesonic state, indicating that no matter the initial pion population, only the stable charged state should survive.
We have as well discussed the Pauli blocking effect and its relevance for leptonic decays. These decay show an interesting nonmonotonic behavior as a function of the isospin chemical potential. Finally, we have briefly commented on the $\pi_0 \to \gamma \gamma$ decay, that is not expected to be stronlgy suppressed. 

We have not studied the decay channels of the kaons, although they are certainly interesting. In particular, it should be intriguing to study the $K\text{-short} \to 2 \pi$ decay, because both the initial and final states are strongly dependent on the values of $\mu_I$ and $\mu_S$. Regarding the charged kaons, we observe that given that the $\tilde K^+$ state is massless in the kaon condensed phase and can be very light in the pion condensed phase, it should be possible for charged pions to decay in  $\tilde K^+$. Regarding the $\tilde K^-$,  it is perhaps more interesting to study the leptonic decays.
We note that for large values of $\mu_S$, see the bottom panel of Fig.~\ref{fig:masses}, the $K^-$ approaches the $\tau$ lepton mass. However, in our model it does not  become heavier than the $\tau$ lepton, thus the   $K^-\to \tau \bar\nu_\tau$ channel should remain closed. 

\acknowledgements{We are grateful to R.~Casalbuoni and to P.~Colangelo for stimulating discussion on  the topics discussed in the present paper.\\
The research of A.M. is supported by Progetto Speciale Multiasse "Sistema Sapere e Crescita"
PO FSE Abruzzo 2007 - 2013}

\begin{appendix}
\section{$\mathcal{L}_{int}$ and amplitudes in the condensed phases} \label{appA}
As shown in \cite{Scherer:2002tk}, the interaction Lagrangian relevant for pion decay is 
\be
\mathcal{L}_{int}=-i\frac{F_0^2}{2} \Tr\[ l_\m \Sigma^\dag \p^\m \Sigma \]\,,
\ee
with the left handed current given by
\be
l_\m=-\frac{g}{\sqrt{2}}(W_\m^+T_++W_\m^- T_-)\,,
\ee where
\be
T_+=\( \begin{array}{ccc}
 0 & V_{ud} & V_{us} \\
0 & 0 & 0 \\
0 & 0 & 0 \\
\end{array} \) \quad \text{and}\quad T_-=(T_+)^\dag\,.
\ee
At the leading order in the fields we obtain
\be
\mathcal{L}_{int}=\frac{F_0}{4}\Tr\[l_\m \bar{\Sigma}^\dag(\p^\m\phi)\bar{\Sigma}+l_\m(\p^\m\phi)\] \label{Lint}\,,
\ee
and the vertex factor in the various  phase is obtained substituting the pertinent expression of $\bar{\Sigma}$  in Eq.~\eqref{Lint}. In the normal phase $\bar{\Sigma}=\bar{\Sigma}_N$, and we obtain the well known result for $\pi$-$W$ boson interaction, see for example~\cite{Scherer:2002tk}.
For the vertex factor in the pion condensation phase we have to use $\bar{\Sigma}_\pi$ given in Eq.~\eqref{pion}, and at the leading order in the mesonic fields we obtain
\begin{widetext}
\be
\mathcal{L}_{int}^{\phi_1\phi_2}=-\frac{F_0 g V_{ud}}{4\sqrt{2}} \{ \p^\m \phi_1 \[2 \cos^2(\a) (W_\m^+ +W_\m^-)\]+\p^\m \phi_2 \[2 i (W_\m^+-W_\m^-)\] \}\,,
\ee
\end{widetext}
that is still written in terms of the unrotated mesonic states. Using the rotated fields defined in Eq.~\eqref{eq:inverted}  we readily obtain
\begin{widetext}
\be
\mathcal{L}_{int}^{\tilde{\pi}^+\tilde{\pi}^-}(k)=-\frac{F_0 g V_{ud}}{2\sqrt{2}} [ \cos \alpha_K(-i k^\m)(U_{11}^* \tilde{a}_+ +U_{21}^*\tilde{a}_-)((\e_\m^+)^*+(\e_\m^-)^*)+ (-i k^\m)(U_{12}^* \tilde{a}_+ +U_{22}^*\tilde{a}_-)(i(\e_\m^+)^*-i(\e_\m^-)^*) ]
\ee
\end{widetext}
where the $U$-matrix elements are given in Eq.~\eqref{U}, $\epsilon_\mu$ is the polarization of the $W^\mu$ boson and  $ \tilde{a}_-$ and $ \tilde{a}_+$ are the destruction operator of the $\tilde{\pi}^-$ and $\tilde{\pi}^+$ fields, respectively. 
From this expression we can handily read the vertex factors of Eq.~\eqref{eq:vert1} and~\eqref{eq:vert2}.

\end{appendix}

\bibliographystyle{ieeetr}
%\bibliographystyle{h-physrev}
%\bibliography{BIB}
%\end{document}

\end{document}